\documentclass[11pt,twoside]{article}
\usepackage{asp2010}
\usepackage{natbib}
\usepackage{units}

\allowtitlefootnote

\newcommand{\mockalph}[1]{}

\aspvolume{471} % Insert volume number from ASPCS
\aspvoltitle{Origins of the Expanding Universe: 1912-1932}
\aspcpryear{2013} % Insert year received from ASPCS
\aspvolauthor{Michael J. Way and Deidre Hunter, eds.}

\def\japblank{}

\begin{document}

\title{``The Waters I am Entering No One yet Has Crossed'': Alexander Friedman
and the Origins of Modern Cosmology\footnotemark}
\titlefootnote{A shorter version of this paper appeared in \cite{belenkiy:38}}

\author{Ari Belenkiy\affil{Simon Fraser University, Department of Statistics and Actuarial Science, BC, Canada}}

\begin{abstract}
Ninety years ago, in 1922, Alexander Friedman (1888-1925) demonstrated
for the first time that the General Relativity equations admit non-static
solutions and thus the Universe may expand, contract, collapse, and even be
born. The fundamental equations he derived still provide the basis for the
current cosmological theories of the Big Bang and the Accelerating Universe.
Later, in 1924, he was the first to realize that General Relativity allows the
Universe to be infinite. Friedman's ideas initially met strong resistance from
Einstein, yet from 1931 he became their staunchest supporter. This essay
connects Friedman's cosmological ideas with the 1998-2004 results of the
astronomical observations that led to the 2011 Nobel Prize in Physics. It also
describes Friedman's little known topological ideas of how to check General
Relativity in practice and compares his contributions to those of Georges
Lema\^{i}tre. Recently discovered corpus of Friedman's writings in the Ehrenfest
Archives at Leiden University sheds some new light on the circumstances
surrounding his 1922 work and his relations with Paul Ehrenfest.
\end{abstract}

\hfil ``L\'acqua ch\'io prendo giammai non si corse.'' \par
\hfil Dante, Paradiso Canto II

\section{Introduction}
The 2011 Nobel Prize in Physics was assigned to scientists who
independently confirmed that the Universe presently expands in an accelerating
manner. Thus one of the scenarios described by Alexander Friedman in 1922 and
1924 was recognized as true. Although the essay ``Accelerating Universe''
\citep{nobel2011} composed by the Class for Physics of the Swedish Royal Academy of
Science to elucidate the ``scientific background of the Nobel Prize in Physics
2011'' cites both of Friedman's works,
\cite{1922ZtesPhy...10..377F,1924ZPhy...21..326F},
regrettably, in the essay text, Friedman's contribution is distorted. It
mistakenly ascribes to \cite{1922ZtesPhy...10..377F} the discovery
``that Einstein's steady state (sic!) solution was really
unstable.''\footnote{This fact was discovered by \cite{1930MNRAS..90..668E}.}
Then it erroneously asserts ``in 1924, Friedman presented his full equations.''
Finally, it wrongly states ``in 1927, the Belgian priest and physicist Georges
Lema\^{i}tre working independently from Friedman performed similar calculations
based on General Relativity (GR) and arrived at the same results.''

This essay remedies these errors. The facts are: already in 1922
\cite{1922ZtesPhy...10..377F} had set the correct framework for GR,
suggesting the most general ``line element'' for the positively
curved space, derived the set of correct equations (now the ``Friedman
equations''), solved them and discussed all three major scenarios for the
expanding Universe. As well, he introduced the expression ``Expanding Universe''
(in his words: ``The Monotone World''). \cite{1924ZPhy...21..326F}
further revolutionized the discourse on GR presenting the idea of an infinite
Universe, static or non-static, with a constant negative curvature, completing
what would be later known as the ``FLRW'' metric.  

Most of the biographical details can be found in
\cite{2006alfr.book.....T} or \cite{friedman1966works}.
However, a recent discovery of a number of Friedman's papers and letters in the
Ehrenfest Archives at Leiden University sheds new light on some particular
circumstances surrounding the discoveries of the Petrograd physicist. The exact
references to the works cited in the text of this essay, as well as the details
of GR theory, can be found in
\cite{2009deu..book.....N}. The translation of several excerpts from 
\cite{friedman1966works} from the Russian is this author's.

\section{Alexander Friedman: A short but very accomplished life}

Born in 1888 and raised in St. Petersburg, Friedman studied mathematics
at St. Petersburg University under the guidance of Vladimir Steklov, in
parallel attending Paul Ehrenfest's physics seminars. Upon graduation in 1910,
he worked primarily in mathematical physics and its applications in meteorology
and aerodynamics. From the outbreak of World War I, Friedman served with the
Russian air force at the Austrian front as an instructor in ballistics. Taking
part in several air reconnaissance flights, he was awarded the military cross
for his courage. 

When, after the February 1917 Revolution, dozens of new universities
were established across Russia, on Steklov's recommendation
Friedman obtained his first professorship in mechanics in Perm near the
Ural Mountains. The faculty included several eminent mathematicians. During
the Russian Civil War, Perm changed hands twice and the teaching
conditions were miserable; the science library was practically missing, as
Friedman often complained in the letters to Steklov. 

 With the Civil War coming to a close, in 1920, Friedman returned to
his \emph{alma mater}, now Petrograd, and started working as a physicist
at the Main Geophysical Observatory, rising to director by 1925. He also
lectured on mechanics at the Polytechnic Institute and the Railway Institute.
Most of his personal research at this time was oriented toward the theories of
turbulence and aerodynamics. Additional professional commitments occasioned
Friedman's parallel investigations into Niels Bohr's quantum theory and Albert
Einstein's general relativity. A month before his untimely death from typhus in
September 1925, Friedman made a record breaking 7,400-meter air balloon flight
risking his life to conduct health-related scientific experiments. His
recollections of this flight were published
posthumously \citep[][pp. 382-5]{friedman1966works}. 

\begin{figure}
\centering
\includegraphics[scale=0.4]{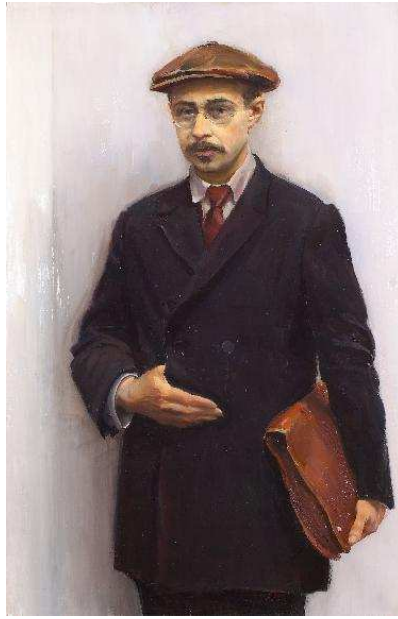}
\caption{Portrait by M.M. Devyatov. Alexander Friedman. Petrograd, 1925.
Courtesy of the Voeikov Main Geophysical Observatory
(St. Petersburg), http://www.voeikovmgo.ru/ru/istoriya.}\label{belenkiyfig01}
\end{figure}

Einstein's special theory of relativity was well-known in
Russia from its inception in 1905, but awareness of GR,
published in 1915, was delayed due to the First World War. The news of GR, along
with the results of the British 1919 astronomical expedition led by Arthur
Eddington and confirming GR's prediction for the gravitational bending of light,
caused tremendous excitement in both the scientific milieu and general public
throughout revolutionary Russia, where it was considered as
another \emph{revolution} -- though in science. Finally, in 1921, shipments of European
scientific publications resumed providing Russian scientists with sufficient
access to the contemporary scientific literature. Physicist Vsevolod Frederiks,
on his return to Petrograd in 1920, brought insider's information: interned in
Germany during the war, he worked at G\"{o}ttingen
University as a private assistant to David Hilbert, who wrote GR's
equations in covariant form in 1916, about the same time as
Einstein.\footnote{The exact timing remains somewhat controversial, see
\cite{Corry1997}.}

In collaboration with Frederiks, Friedman organized a seminar
dedicated to the study of GR. Together they aimed to write a comprehensive
textbook on GR; the first volume, devoted to tensor calculus, appeared in 1924.
In parallel, in his own book,
\emph{The World as Space and Time} \citep{friedman1923mir}
Friedman developed a philosophical interpretation of GR. But his fame rests on
two papers, published in \emph{Zeitschrift f\"{u}r Physik} in 1922 and 1924
\citep{1922ZtesPhy...10..377F,1924ZPhy...21..326F},\footnote{There also exist
English translations of both
papers \citep{1999GReGr..31.1991F,1999GReGr..31.2001F}.} with
new solutions of GR equations. In these papers he introduced the fundamental
idea of modern cosmology -- that the Universe is dynamic and may evolve in
different manners, for example, starting from singularity.

\section{Cosmology before Friedman: Rivalry between two static Universes' models}

The 16 (or actually 10 different) equations of GR are: 
\begin{equation}\label{belenkiyeq1} 
R_{ik} -\frac{1}{2} g_{ik} \overline{R}-\Lambda g_{ik} =-\kappa T_{ik} ,        
\end{equation}

\noindent where indexes $i$ and $k$ run from 1 to 4. The first three indexes
relate to space while index 4 relates to time, $g_{ik}$ is the metric tensor, $R_{ik}$
is the Ricci tensor representing 2-dimensional curvatures, $\overline{R}$ is the
scalar space-time curvature, constant
$\kappa = 8\pi \japblank G/c^{2}=1.87\times10^{-27}$ cm $g^{-1}$. $G$ is the
gravitational constant, $c$ is the speed of light in vacuum, and $T_{ik}$ is the
energy-matter tensor representing the ``inertia'' of the world. The latter was
assumed to be $T_{11} =T_{22} =T_{33} =-p$, where $p$ is the pressure of
radiation, $T_{44} =c^{2} \rho \japblank g_{44} $, where $\rho$ is average density of
matter in the Universe, and $T_{ik} =0$ for non-diagonal elements. Einstein
readily considered a simplification with $p=0$. The equality sign in
Equation \ref{belenkiyeq1} signifies the ``equivalence principle" between gravity (on
the left) and inertia (on the right).\footnote{There is a variant opinion
that the \emph{equivalence principle} means ``gravity $=$ space-time'' but see 
\citet[][p. 76]{1920stga.book.....E}.}

The left side of Equation \ref{belenkiyeq1} is highly non-linear in $g_{ik}$
and its derivatives of the first and second order. To assure stability of the
solution, Einstein introduced a linear term, $\Lambda g_{ik}$, where the
coefficient $\Lambda$ became known as the ``cosmological constant.''\footnote{Since the
metric tensor $g_{ik}$  has dimension $s^{2}$  and the equations (1)
are dimensionless, $\Lambda$  has dimension $s^{-2}$.}

Since finding a solution to this system of equations requires great
ingenuity, only two simple solutions were discovered by 1922, one by Einstein,
and the other by the Dutch astronomer Willem de Sitter.

The so-called ``solution A,'' found by \cite{1917SPAW.......142E}, represented a
spatially 3-dimensional spherical, finite Universe with curvature radius R
constant in space and time. In coordinates $(\chi ,\theta ,\varphi )$ the
elements of metric tensor are: 
\begin{equation}\label{belenkiyeq2}      
g_{11} =-\frac{R^{2} }{c^{2} } ,\; g_{22} =-\frac{R^{2} }{c^{2} } \sin
^{2} \chi ,\; g_{33} =-\frac{R^{2} }{c^{2} } \sin ^{2} \chi \japblank \sin ^{2} \theta
,\; g_{44} =1,
\end{equation}
while all other $g_{ik} =0$. 
Einstein's Universe is a 3-dimensional sphere with a fixed radius,
which evolves in time as a 4-dimensional cylinder. 

Applying GR equations \ref{belenkiyeq1}, the only solution comes
when the two (initially independent) parameters, $\Lambda$ and $\rho$, become interconnected
and expressed via radius R:
\begin{equation} \label{belenkiyeq3}
\Lambda=\frac{c^{2} }{R^{2} } \ \quad \& \ \quad \rho =\frac{2}{\kappa \japblank R^{2} } .
\end{equation}
Multiplying $\rho$ by the volume of the 3-dimensional sphere,
$V=2\pi ^{2} R^{3} $, the Universe's mass could be found as
$M=4\pi^{2} R/\kappa$.

The remarkable consequence of ``solution A'' was that a good estimate
of average density leads to an estimate of the radius and mass of the Universe.
With estimate $\rho =2\times10^{-27}$ g cm$^{-3}$, suggested
by \cite{deSitter1917MNRAS..78....3D},
it seemed that Einstein had achieved his goal and the ``final theory'' of the
spherical Universe was constructed, with constant radius $R=750\times 10^{24}$ cm,
or 800 Mly.\footnote{We use ``Mly'' for ``million light years''
and ``Gly'' for ``billion light years.''}  De Sitter's discovery of another
solution a month later came as a cold shower for Einstein.

The so-called ``solution B,'' found by \cite{deSitter1917MNRAS..78....3D},
presented a different Universe. Though the rest of $g_{ik}$ were the
same as in Equation \ref{belenkiyeq2}, importantly $g_{44} =\cos ^{2} \chi $,
i.e., time was curved in the direction of the ``radial'' spatial
coordinate $\chi $. Though spatially spherical, de Sitter's Universe has a point
different from any other point, a \emph{center}, whereas in Einstein's
solution every point is equivalent to any other. Rays of light don't move along
the space geodesics in de Sitter's Universe, with the exception of those which
pass through the center.

To satisfy GR equations \ref{belenkiyeq1} this solution necessitates
a non-zero cosmological constant but zero density:
\begin{equation}\label{belenkiyeq4}
\Lambda =\frac{3c^{2} }{R^{2} } \ \quad \& \ \quad \rho =0.
\end{equation}
The major feature of this model, absence of matter $(M=\rho \japblank V=0)$,
violated Ernst Mach's principle that ``inertia cannot exist without
matter'' and thus made this solution unacceptable for Einstein. Moreover, the
function $\cos ^{2} \chi $ before the time component suggests a singularity
at $\chi = {\pi }/{2} $. This mystical locus, where time ``stops to flow,''
Hermann \cite{weyl1918raum} called the ``horizon.'' The space of every
observer was surrounded by such a ``horizon'' though the latter was unreachable.
However, as de Sitter noticed, the alleged ``slowing of time'' along the radial
component c provided the means to explain the shifts $z=\delta \lambda /\lambda$
of absorption lines in the spectra of nebulae,\footnote{The nebulae were not
understood as distinct \emph{island universes} until Hubble's work on
Cepheids in M31 and M33 in 1924 \citep{Hubble1925PA.....33..252H}.}
first observed by Vesto Slipher at Lowell Observatory in Flagstaff, Arizona, in
1912 \citep{Slipher1913LowOB...2...56S}.
Saying that ``the observations are still very uncertain, and conclusions
drawn from them are liable to be premature,'' de Sitter discussed only three
nebulae, whose radial velocities have been determined ``by more than one
astronomer.'' The spectrum of the Andromeda nebula showed a blue-shift equivalent
to a speed of 311 km s$^{-1}$, but the other two showed more
pronounced redshifts of 925 and 1,185 km s$^{-1}$.

Computing the average of three velocities, de Sitter related the
``average'' redshift of 600 km s$^{-1}$ to his ``solution B'' via the
formula $z=(1/2)\sin ^{2} \chi $, but deduced from it an absurdly small
Universe's curvature radius $R$ as $5\times10^{24}$ cm or 4.5 Mly.
Indeed, the 100-inch telescope at the Mount Wilson observatory, established in 1917,
could reach as far as 150 Mly.\footnote{According to \citet{1927ASSB...47...49L,
1931MNRAS..91..483L}: ``The range of the 100-inch Mount Wilson telescope
is estimated by Hubble to be 5$\times$10$^{7}$ parsec."}

\begin{figure}
\centering
\includegraphics[scale=0.9]{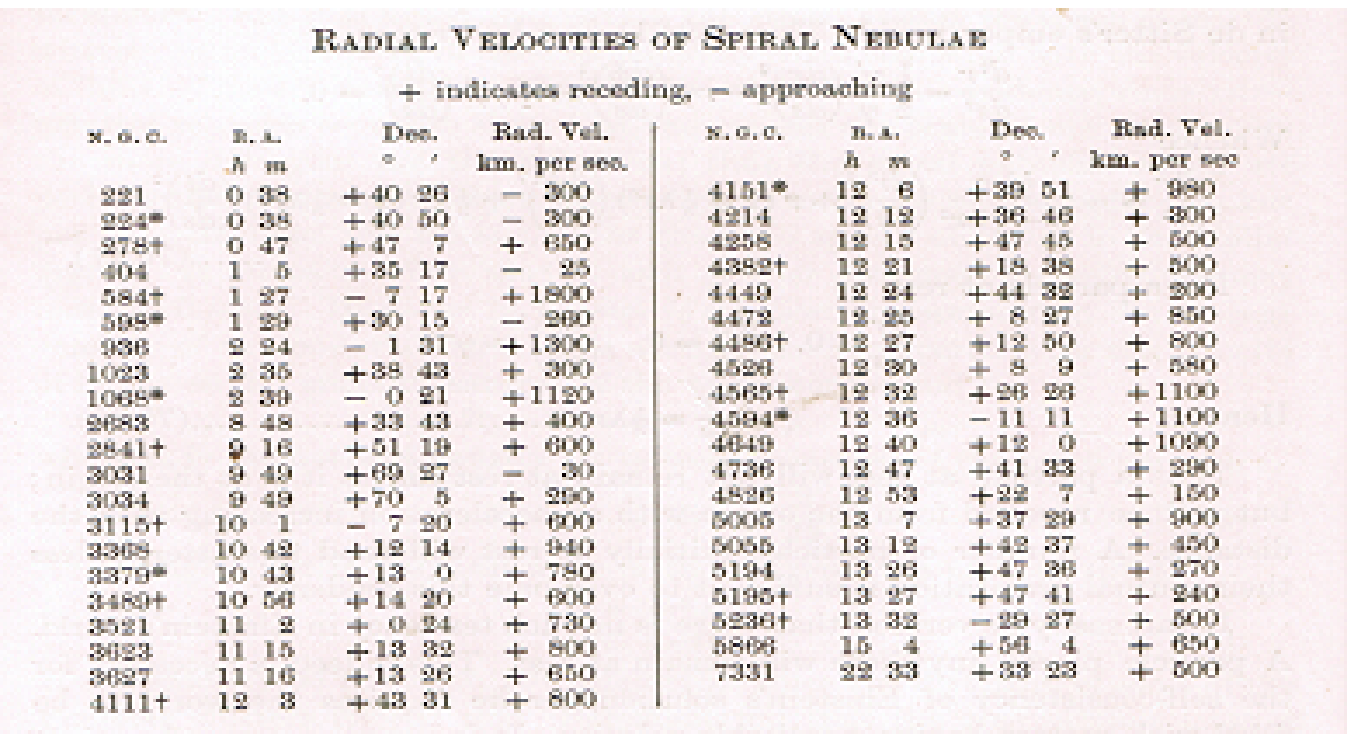}
\caption{``Radial velocities'' of 41 spiral galaxies (then
known as ``nebulae'') found by Vesto Slipher. 36 galaxies have positive
velocities while only 5 negative. Each galaxy is identified by its NGC catalog
number and its equatorial coordinates (Right Ascension and Declination). From
\citet[][p. 162]{1923mtr..book.....E}.} \label{belenkiyfig02}
\end{figure}

Meanwhile the evidence for the redshifts was mounting mainly due to
Slipher's efforts, and by 1923 reached a score of 36 among 41 spiral nebulae.
Eddington popularized this fact in \cite{1923mtr..book.....E} (see Fig. \ref{belenkiyfig02}).

Ready to identify the redshifts with the Doppler effect, most of the
workers in this field adopted ``solution B'' as another means to test GR,
looking for a better formula for the redshift and hoping to dismiss the weird
``horizon'' using various coordinate transformations. The first goal was
achieved by Hermann \cite{weyl1923PZ} and later by Ludwik
\citet[][p. 523]{silberstein1924theory}, who improved the formula for the
redshift to $z=\pm \sin \chi +\sin ^{2} \chi $ and then deduced the value of
the radius $R$ ranging from 80 to 120 Mly. The second goal was achieved by Kornel
\cite{1923ZPhy...17..168L}, who reworked ``solution B'' into a non-static solution
with an exponentially growing radius. Later Georges
\cite{Lemaitre1925JMP} quite elegantly repeated both results. But by then, an
absolutely novel and daring idea had been born and developed by an outsider, a
physicist from far away ``revolutionary'' Petrograd.

\section{Friedman's Expanding Universes: Three major scenarios}

\begin{figure}[ht]
\centering
\includegraphics[scale=0.58]{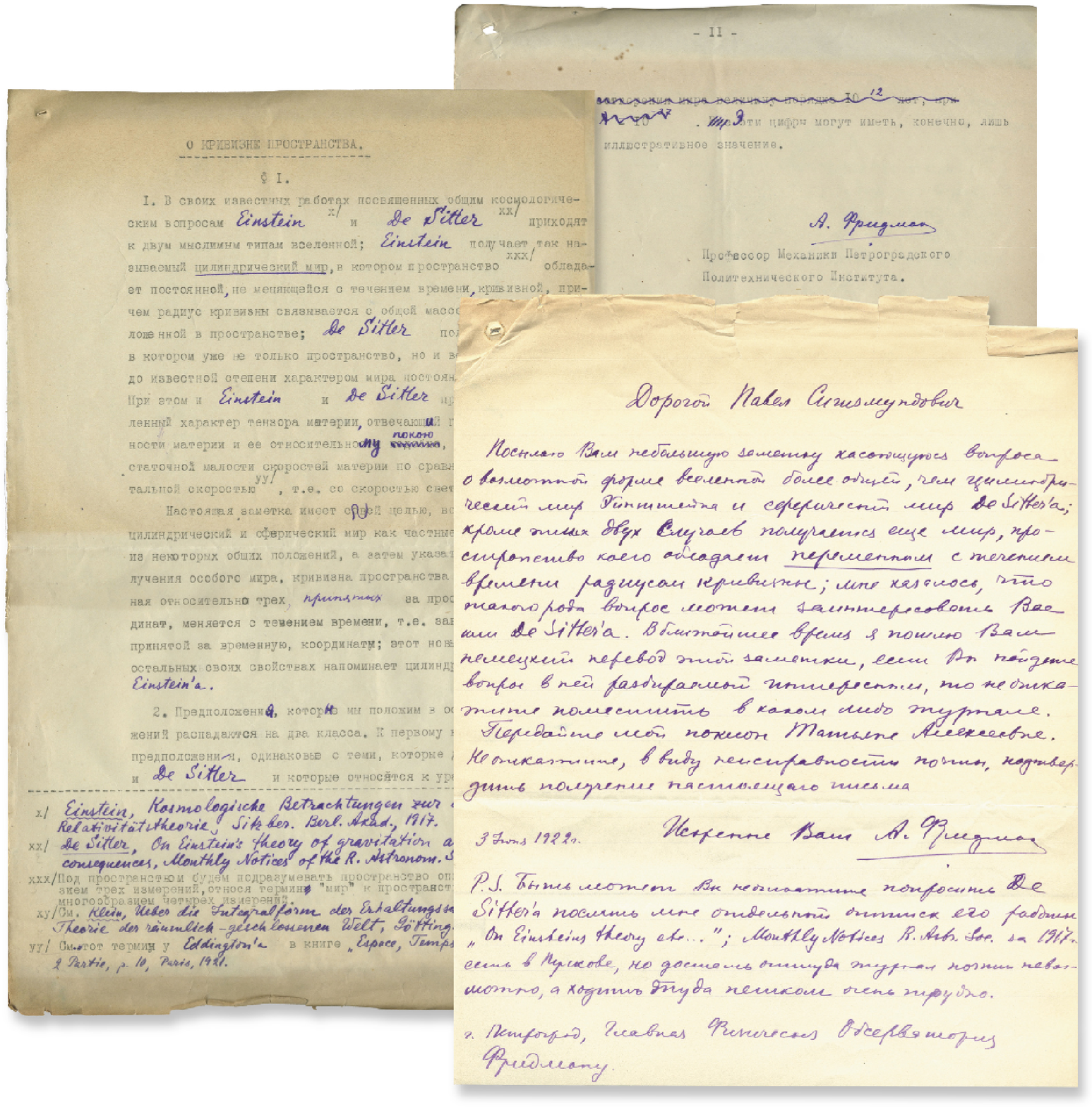}
\caption{The first and last pages of Alexander Friedman's
original draft ``On the Curvature of Space'' (in Russian) sent to Paul Ehrenfest
on June 3, 2012, accompanied by a letter, for possible submission to
\emph{Zeitschrift f\"{u}r Physik}. Note several crossed lines on the top of the
last page. Courtesy of Lorentz-Institute, Leiden University
\citep{Beenakker2012}.} \label{belenkiyfig03}
\end{figure}

On June 29, 1922, \emph{Zeitschrift f\"{u}r Physik} accepted the paper
``On the Curvature of Space'' by A. Friedman of Petersburg, submitted to the journal by
Paul Ehrenfest. Though \cite{1922ZtesPhy...10..377F} cites there
the original works by Einstein (1917) and
\cite{deSitter1917MNRAS..78....3D}, he certainly learned of
these works from Arthur Eddington's \emph{Space, Time
and Gravitation} \citep{1920stga.book.....E}, available to him in the French
edition of 1921, with deliberations on the worth of the two models.\footnote{This
can be inferred from his letter to Paul Ehrenfest of June 3, 1922
(Fig. \ref{belenkiyfig03}).}  But instead of taking sides, the
scientist from Petrograd approaches the problem from a wider view point. 

The physical demand of homogeneity and isotropy of space does not
necessitate a static Universe. Focusing on the most general form of spherical
metric, \cite{1922ZtesPhy...10..377F} finds on top of static
solutions A and B, a new class of non-static solutions of GR
equations \ref{belenkiyeq1}. 

Friedman's dynamical solution is a generalization of Einstein's 3-dimensional hypersphere
Eq. \ref{belenkiyeq2} with a constant in space but changing in time curvature
radius $R(t)$. In this case equations \ref{belenkiyeq1} yield two
ordinary differential equations for $R(t)$ (the ``Friedman equations'' in
modern terminology):\footnote{The ``dot'' and two ``dots'' above the variable
denote, as usual, the first and second time derivatives.}
\begin{equation}\label{eq:F1}
\frac{2\ddot{R}}{R} +\frac{\dot{R}^{2} }{R^{2} } +\frac{c^{2} }{R^{2}
} -\Lambda =0                                                 
\end{equation}
and
\begin{equation}\label{eq:F2}
\frac{3\dot{R}^{2} }{R^{2} } +\frac{3c^{2} }{R^{2} } -\Lambda =\kappa c^{2} \rho .
\end{equation}
The second-order Ordinary Differential Equation (Eq. \ref{eq:F1})
Friedman integrates directly, without using
Bianchi identities, arriving at the fundamental equation that governs the
dynamics of the Universe:
\begin{equation} \label{belenkiyeq5}
\frac{1}{c^{2} } \dot{R}^{2} =\frac{A-R+\frac{\Lambda }{3c^{2} } R^{3} }{R}    .
\end{equation}
Comparing Eq. \ref{belenkiyeq5} with Eq. \ref{eq:F2}
Friedman finds that the constant of integration, $A$, is related to the
density $\rho$ as
\begin{equation}\label{belenkiyeq6}
\rho=\frac{3A}{\kappa R^{3} }. 
\end{equation}
Multiplying Eq. \ref{belenkiyeq6} by the volume of hypersphere
$V=2\pi^{2} R^{3}$, Friedman obtains $A=\kappa M/6\pi ^{2}$. Thus
$A$ is proportional to the Universe's mass \emph{M} with
constant $\kappa$ and represents the \emph{gravitational radius} of the Universe.

The rest of the 1922 paper is dedicated to analysis of
Eq. \ref{belenkiyeq5}, which after integration becomes
\begin{equation}\label{belenkiyeq7}
t=\frac{1}{c} \int _{R_{0}
}^{R}\sqrt{\frac{x}{A-x+\frac{\Lambda }{3c^{2} } x^{3} } } dx+t_{0} .
\end{equation}
Taking $R_{0}$ to be the present radius of the Universe,
$t_{0}$ designates, in Friedman's words, ``the time that passed from
Creation,'' or \emph{the age of the Universe}. 

\begin{figure}
\centering
\begin{tabular}{ccc}
\includegraphics[scale=0.35]{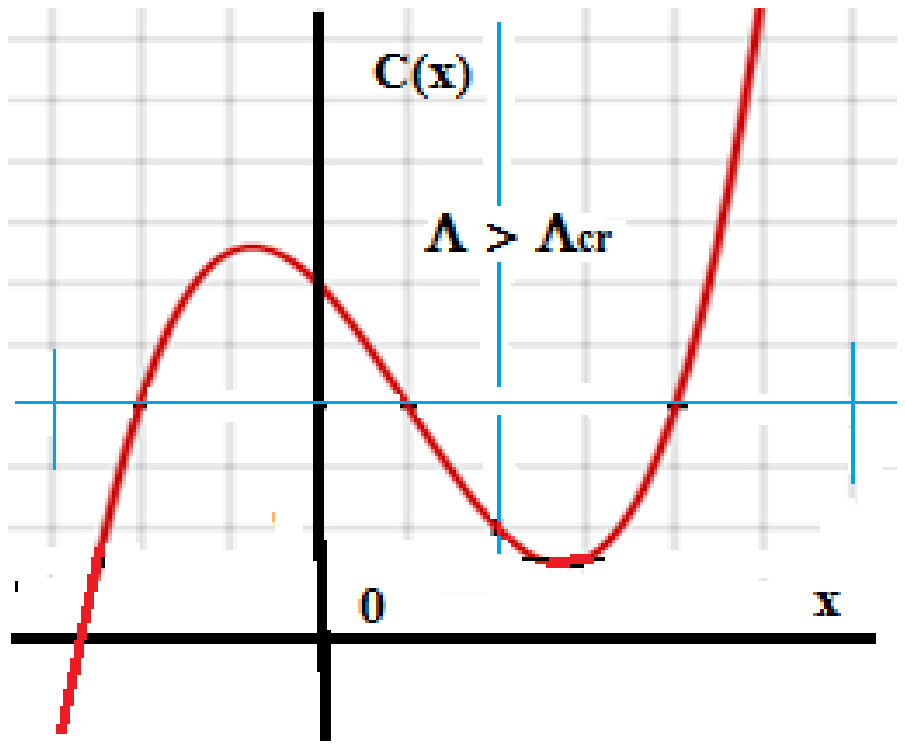} &
\includegraphics[scale=0.39]{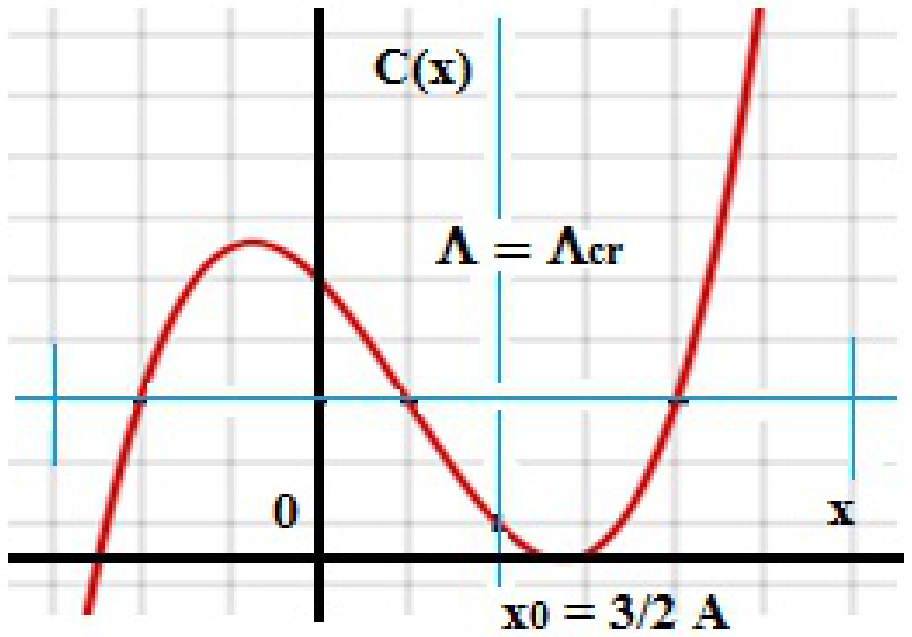} &
\includegraphics[scale=0.41]{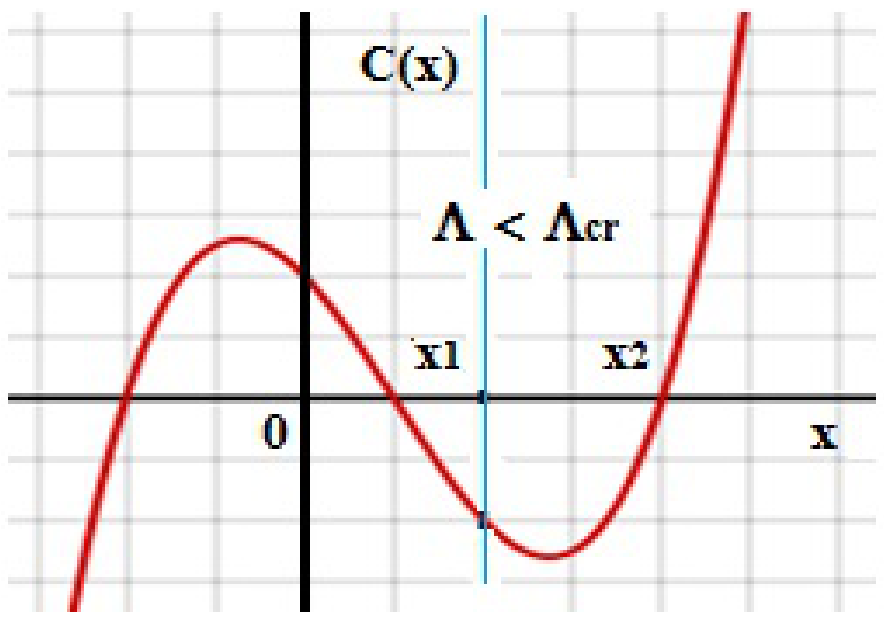} \\
\end{tabular}
\caption{Friedman's fundamental cubic $C(x)$ that governs the
dynamics of the Universe. Its three coefficients are three fundamental
constants: cosmological constant, the sign of curvature of space and the
gravitational radius of the Universe. Depending on their values, the cubic may
have either a. 0, or b. 1 double or c. 2 simple positive roots, which define six
scenarios of the Universe's evolution.}\label{belenkiyfig04}
\end{figure}

The right hand-side of Eq. \ref{belenkiyeq7} has physical meaning only
when the cubic $C(x)$ in denominator is \emph{positive}. The cubic $C(x)$ can
be positive in three ways (Fig. \ref{belenkiyfig04}): on a semi-infinite
interval that may start either 1)
at 0 or 2) at a positive real number, or 3) on a finite segment. This
defines three major scenarios of the Universe's evolution
(Fig. \ref{belenkiyfig05}). If $C(x)$ has a double positive root, then three
more scenarios are possible (Fig. \ref{belenkiyfig06}). 

\begin{enumerate}
\item  The first scenario comes if the cubic has no positive
roots and thus is positive on $(0,\infty)$. This happens
when $\Lambda > {4c^{2} }/{9A^{2} } $, i.e., $\Lambda$ is positive
and larger than a certain critical value for a given density $\rho$.
The Universe starts from singularity $R=0$ at $t=0$; its asymptotic behavior at
infinity is $R\approx R_{0} \exp[\sqrt{\Lambda/3}~(t-t_{0} )]$. At some
inflexion point its expansion changes from deceleration to acceleration.
According to Eqns. \ref{belenkiyeq5} and \ref{belenkiyeq7}, the inflexion point
occurs where expression $C(x)/x$ reaches its minimum, i.e., when the radius of
Universe reaches
\begin{equation}\label{belenkiyeq8}
R_{f} =\left(\frac{3c^{2}
A}{2\Lambda } \right)^{\frac{1}{3} } =\left(\frac{\kappa \japblank c^{2} \japblank \rho
}{2\Lambda } \right)^{\frac{1}{3} } R_{0}   .
\end{equation}
Friedman called this scenario ``The Monotone World of the first kind''
(``M 1'' in Fig. \ref{belenkiyfig05}). 

\begin{figure}[ht]
\centering
\includegraphics[scale=0.7]{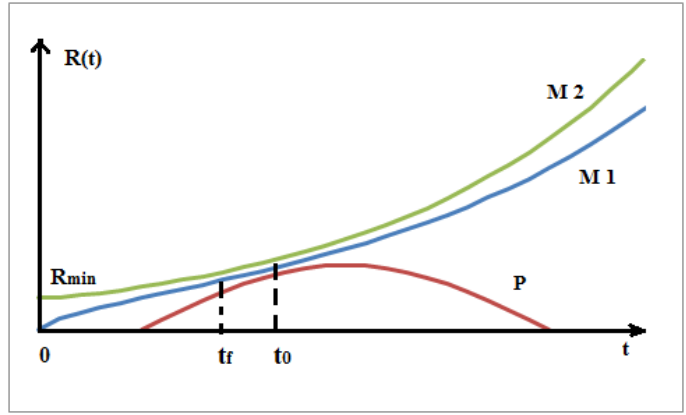}
\caption{Three possible major scenarios of the Universe's
evolution according to \cite{1922ZtesPhy...10..377F}. The M1 world
shows expansion from singularity with a flex point that signifies existence of
two stages of evolution: deceleration and acceleration.
\citet[][p. 84]{1961cosm.book.....B} calls it ``Lema\^{i}tre's case.''
The M2 world shows
expansion from the non-zero radius to infinity. The P world shows periodic
evolution, with expansion and contraction phases and point of maximum radius in
between. Point $t_{0}$ is the current stage of the Universe and point $t_{f}$
is the inflexion point in the M1 world.}\label{belenkiyfig05}
\end{figure}

\item The second situation occurs when $0<\Lambda<
{4c^{2} }/{9A^{2} } $. In this case the cubic has two positive roots,
$x_{1}$ and $x_{2}$ and is positive in intervals $(0,x_{1})$
and $(x_{2} ,\infty)$. This presupposes two different scenarios: 2a
and 2b. In the 2a scenario, expansion oscillates between $R=0$ and $R=x_{1}$.
This is the ``periodic'' solution, viable also for a wider range of $\Lambda$'s,
discussed next. In the 2b scenario, expansion starts from a non-zero radius,
$R_{\min}$, equal to the greater root, $x_{2}$, and continues
forever in accelerating mode. Its asymptotic behavior at infinity again is
$R\approx R_{\min } \exp[\sqrt{\Lambda/3}~(t-t_{0}) ]$. Friedman named
this scenario
``The Monotone World of the second kind'' (``M 2'' in Fig. \ref{belenkiyfig05}). 

Taking in Eq. \ref{belenkiyeq7} $A=0$ or, equivalently, $\rho=0$ (de Sitter's case),
this solution simplifies to
$R=\varsigma \japblank \cosh ({c\japblank t}/{\varsigma })$,
where $\varsigma =c\japblank \sqrt{3/\Lambda }$ has
meaning of the minimal radius, $R_{min}$, while $\Lambda$ accepts de Sitter's value
$\Lambda ={3c^{2} }/{R_{\min }^{2} } $, as in Eq. \ref{belenkiyeq4}. Since
the function $\cosh ({c\japblank t}/{\varsigma })$ smoothly continues to $t<0$, a
symmetric (left) branch may be added to the M2 curve in Fig. \ref{belenkiyfig05}.
This change would
modify the M2 scenario: the first phase of the Universe's evolution becomes
infinitely long contraction to $R_{\min } =\varsigma $ , followed by an
infinitely long expansion back to infinity.
Note the intermediate case, $\lambda _{cr} ={4c^{2}}/{9A^{2}} $, 
where the cubic has a positive double root. This case was chosen
later by \cite{1927ASSB...47...49L}.
Friedman considered it a ``limiting case'' -- see below.

\item The third scenario results either from scenario 2a, or when
$\Lambda \le 0$. In both cases the cubic has one positive root, $x_{1}$ (when
$\Lambda =0$ the cubic reduces to the linear function with one positive root)
and the interval of positivity is $(0,x_{1})$.  The Universe starts from
singularity $R=0$ at $t=0$ and expands in a decelerating manner, then stops at
$R=x_{1}$ and begins contracting back into singularity. The life of the
Universe is finite. Friedman called this scenario the ``Periodic World''
(``P'' in Fig. \ref{belenkiyfig05}) and found its approximate period
(for small $\Lambda$) as
\begin{equation}\label{belenkiyeq9}
T_{p} =\frac{2}{c} \int
_{0}^{x1}\sqrt{\frac{x}{A-x+\frac{\Lambda }{3c^{2} } x^{3} } } dx \approx
\frac{\pi \japblank A}{c} =\frac{\kappa \japblank M}{6\pi \japblank c} .
\end{equation}
Upon assuming the mass of the Universe \emph{M} is equal to
$5\times10^{21} $masses of the Sun, Friedman found $T_{p} =10^{10}$
(ten billion) years. 
Unfortunately we cannot repeat this result. Since the sun's mass is
known as $M_{S} =2\times10^{33} g$, we get $M=10^{55} g$ and therefore
\begin{equation}\label{belenkiyeq10}
T_{p} \approx \frac{\pi \japblank A}{c}
=\frac{\kappa \japblank M}{6\japblank \pi \japblank c} =\frac{10^{28} cm}{3\japblank \pi \japblank
c} \approx \frac{10^{27} cm}{c} ,
\end{equation}
which is only 1 billion years.  One can only guess where the error crawled
into Friedman's reasoning.

Considering in Eq. \ref{belenkiyeq7} $A=0$ or, equivalently, $\rho=0$
(de Sitter's case) Friedman's formula for $\Lambda <0$\emph{ }simplifies to
$R=\varsigma \japblank \sin ({c\japblank t}/{\varsigma }) $, where $\varsigma =c\japblank
\sqrt{3/\Lambda } $ plays a role of maximal radius $R_{max}$, while $\Lambda$
accepts de Sitter's value  $\Lambda ={3c^{2} }/{R_{\max }^{2} } $, as in
Eq. \ref{belenkiyeq4}.

\begin{figure}[ht]
\centering
\includegraphics[scale=0.7]{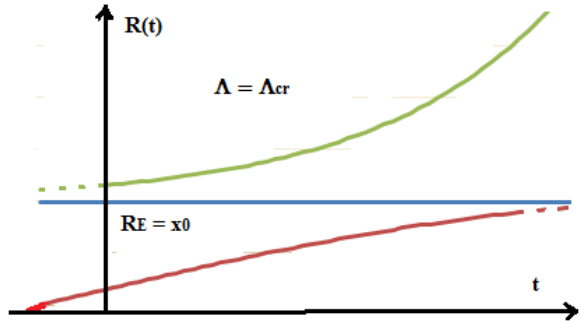}
\caption{Two special cases of Friedman's Universe 
\citep{1922ZtesPhy...10..377F}, both with ``logarithmic infinity.'' The upper
curve is the limiting case of the ``M 2'' world with infinite time elapsed from
the beginning; \citet[][p. 84]{1961cosm.book.....B} calls it
``Eddington-Lema\^{i}tre'' case. The lower curve is a limiting case of the
``Periodic World'' with infinitely long expansion toward the finite radius. The
horizontal line stands for Einstein's ``Solution A.''}\label{belenkiyfig06}
\end{figure}

\item In addition to the three major scenarios,
\cite{1922ZtesPhy...10..377F} mentions two special (``limiting'') cases, which
occur when $\Lambda$ is equal to
$\Lambda _{cr} = {4c^{2} }/{9A^{2} }$
and the cubic is ``degenerate,'' i.e., has a double positive root at $x=3A/2$. 
The double root leads to a logarithmic singularity at the finite
radius $R_{E} =3A/2$. The 2a scenario degenerates into an infinitely long
expansion from singularity \emph{R=0} to the finite radius $R_{E}$ at
infinity: $R\approx R_{E} -\exp[{-\sqrt{\Lambda}~(t-t_{0} )}]$
(Fig. \ref{belenkiyfig06}, lower curve).
The 2b scenario degenerates into an infinitely long expansion $R\approx
R_{E} +\exp[{\sqrt{\Lambda}~(t-t_{0} )}] $from the finite radius $R_{E}$ in the
past ($t<t_{0}$) via the exponential growth
$R\approx R_{0} \exp[\sqrt{\Lambda/3}~(t-t_{0})] $
in the future ($t>t_{0} )$ (Fig. \ref{belenkiyfig06}, upper curve).
The latter scenario was adopted by \cite{1927ASSB...47...49L}. Both curves
asymptotically converge to radius $R_{E} =3A/2$, which can be viewed as
Einstein's radius for the static Universe. This static Universe of Einstein is
the sixth scenario with
$R_{E} =3A/2={2G\japblank M}/{\pi \japblank c^{2} } ={R_{S} }/{\pi } $,
where $R_{s}$ is the Schwarzschild radius.
\end{enumerate}

\section{Friedman's philosophy of the Big Bang and Expanding Universe}

In Friedman's models the cosmological constant $\Lambda$ is a free
parameter to be determined empirically, whereas in Einstein's and de Sitter's
models $\Lambda$ was strictly linked to the curvature of the Universe. It is
the pair, $\rho$ and $\Lambda$, that determines a true scenario for the
Universe's evolution.

In the 1922 paper Friedman remains silent about details, but in his
book \emph{World as Space and Time}, sent to print on September 5, 1922, and
published the following year, he allows himself to say more. There, in the
section ``Matter and the Structure of the Universe,'' \cite{friedman1923mir}
describes the Big Bang scenario in the following words:

\begin{quote}
 A non-static Universe represents a variety of cases. For example, it
is possible that the radius of curvature constantly increases from a certain
initial value; it is also possible that the radius changes periodically. In the
latter case the Universe compresses into a point (into nothingness), then
increases its radius to a certain value, and then again compresses into a point.
Here one may recall the teaching of Indian philosophy about ``periods of life.''
It also provides an opportunity to speak about the world ``created from
nothingness.'' But all these scenarios must be~considered~as curiosities which
cannot be presently supported by solid astronomical experimental data. So
far it is useless, due to the lack of reliable astronomical data, to cite any
numbers that describe the life of our Universe. Yet if we compute, for the sake
of curiosity, the time when the Universe was created from a point to its present
state, i.e., time that has passed from the ``creation of the world,'' then we
get at number equal to tens of billions of usual years.\footnote{Translated
from \citet[][p. 317]{friedman1966works}.}
\end{quote}

It is interesting that here Friedman mentions only the M2 world and
Periodic world but not the M1 world. The reason seems to lie in his low \emph{a priori}
estimate for the cosmological constant, $\Lambda =10^{-37} s^{-2} $, which is crossed out but
still visible in the Russian manuscript of the 1922 paper (see Fig. \ref{belenkiyfig03}).
His estimate $3A=10^{27}$ cm (see Eq. \ref{belenkiyeq10})
implies $\Lambda _{cr} ={4c^{2} }/{9A^{2} } = 36\times10^{-34} s^{-2}$,
i.e. greater than $\Lambda$ by four orders, thus making the M1 scenario impossible.

The choice of small $\Lambda$ was motivated by his cautious estimate of the
age of the Universe as in the last lines of his Russian draft ``of the order of
$10^{12}$ years,'' which he crossed out in the Russian manuscript at the last moment.
Though this figure is reminiscent of James Jeans' estimate of the age of
the solar system,  $10^{13} $ years, Friedman could not have possibly known it.
Jeans seems to advocate such a high estimate not earlier than 1928, the second
edition of his \emph{Astronomy and Cosmogony}.\footnote{\cite{Jeans1928astronomy}}
At least \cite{1920stga.book.....E}, read by Friedman,
is silent on it. Therefore Friedman must have learned it from another source. 

And indeed, \citet[][p. 163]{1920stga.book.....E} gives his
own estimate of the radius of the Universe, $2\times10^{11} $ parsecs, which is
$6\times10^{11} $ light years and thus of the same order as Friedman's $10^{12}$
light years. Eddington's \emph{ad hoc} estimate comes from comparison of the
radius and ``gravitational radius'' of the electron.\footnote{Though both
radiuses used by Eddington, $7\times10^{-56}$ cm and $2\times10^{-13}$ cm,
are smaller than the currently accepted values by factors 2 and 1.4,
respectively, the net-result for the curvature radius of the Universe is nearly
the same.} This could explain Friedman's
somewhat unclear statement \citep{1922ZtesPhy...10..377F} that ``hopefully
$\Lambda$ may be found from electrodynamical considerations.''

\section{Einstein's reception of Friedman's theory in 1922-1923}

The main ideas of the \cite{1922ZtesPhy...10..377F} paper -- a
dynamic character of the Universe, his metric, his equations, and exhaustive
description of possible scenarios for the real Universe have become fundamentals
in contemporary cosmology. However, in 1922 they were mostly ignored or rejected. 

Friedman's paper appeared in print in July 1922 in volume 10 of
\emph{Zeitschrift f\"{u}r Physik} and was noticed by Einstein. Most likely,
Ehrenfest, Einstein's close friend, called his attention to it.  Einstein's
immediate reaction illustrates how unwelcome the idea of a non-static universe
was. In his view, a normal cosmological theory should uphold the static
character of the Universe. Accordingly, Einstein initially found Friedman's
solution ``suspicious'' and in October 1922 published a short note in volume 11
of \emph{Zeitschrift f\"{u}r Physik} suggesting Friedman's derivation contained a
mathematical error. 

In fact, \cite{Einstein1922ZPhy...11..326E} mistakenly concluded that
Friedman's equations (with $p=0$) imply ${d\rho }/{dt} =0$,  i.e., constancy
of density, and therefore of volume and radius \emph{R} -- in contradiction to
the initial assumption. Instead, the correct derivation gives $R^{-3} 
{d(\rho \japblank R^{3} )}/{dt} =0$, which implies constancy of mass and thus
adequately represents the ``conservation of mass'' law.\footnote{Of course,
considering zero pressure $p$ deprived the formulation of this law of its
most general form found later by \cite{1927ASSB...47...49L} as
$c^{2} \japblank dM+p\japblank dV=0$, correctly interpreting it as
``energy spending for the adiabatic expansion of the Universe.''} 

Learning of Einstein's note, on December 6, 1922 Friedman wrote a
lengthy letter to Einstein, presenting his derivations, but Einstein was already
on his world tour, returning to Berlin in March 1923. Only then could
he have read Friedman's letter \citep{2006alfr.book.....T}.
Later, in May, Yuri Krutkov, Friedman's colleague, met
Einstein twice at Leiden, at Paul Ehrenfest's, and clarified the
confusion. Certainly, the meeting was organized by Ehrenfest, who felt himself
personally responsible for presenting Friedman's paper to \emph{Zeitschrift
f\"{u}r Physik.}

Following this meeting, on May 31, 1923, \cite{Einstein1923ZPhy...16..228E}
sent another short
note to \emph{Zeitschrift f\"{u}r Physik} accepting the mathematical correctness
of Friedman's results. However, he opined that ``the solution has no physical
meaning;'' wisely he crossed this out from the proofs at the last moment.
Einstein was unready to accept the idea of the expanding Universe for 8 more
years.\footnote{In his two notes, Einstein inadvertently introduced the
following problem for future historians of cosmology: he christened Friedman
with two n's in his last name whereas Friedman's 1922 paper was signed with
one ``n.'' It seems Friedman followed the ``advice'' and submitted his 1924
paper with two n's.}

\section{In a quest for an infinite Universe: The Universe with negative curvature} 
Already by 1922 Friedman realized that GR equations
\ref{belenkiyeq1} alone fail to provide the final answer for not only the
kinematics of the real Universe but also its global structure (the shape and
size). The means to choose one solution over another needed to come from
elsewhere, for example, astronomy. The Universe in the shape of the
3-dimensional hypersphere $S^{3}$, for example, admits ``ghosts'' -- double images of
the same object in the sky coming from two opposite directions (the second is
from passing the ``antipodal'' point). To avoid such a misleading phenomenon,
\cite{deSitter1917MNRAS..78....3D}
pioneered the idea that the space of directions on the hypersphere
$S^{3}$ must be considered as the basic cosmology space where each pair of
``antipodal'' points might be viewed as one. A newly arising science of
algebraic topology assured that this space is an orientable manifold and its
metric element is the same as that of $S^{3}$.\footnote{This space was
called the \emph{elliptical space} though now it is better known as \emph{real
projective space} $RP^{3}$. The first constructions of the projective spaces
were given by Felix Klein in 1890 and Henri Poincare in 1900
\citep{klein1928vorlesungen}.} De Sitter's idea certainly gained some
hard currency as \cite{1924ZPhy...21..326F} mentioned it favorably
while \cite{1927ASSB...47...49L}
even adopted it, computing the volume of his Universe as $\pi ^{2} R^{3} $,
i.e., half of the $S^{3}$'s volume. 

\begin{figure}[ht]
\centering
\includegraphics[scale=0.7]{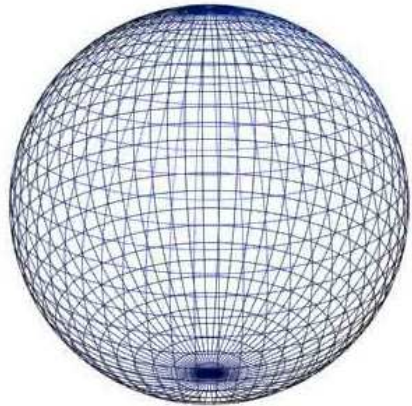}
\caption{A 2-dimensional sphere covered with an  infinite
covering. If one folds the cover perpendicular to one axis then its
two poles remain ``uncovered.''}\label{belenkiyfig07}
\end{figure}

However, Friedman's major concern lay with the notion of the world's
finiteness, which was firmly entrenched in the minds of scientists, largely,
because of Einstein's staunch adherence to Mach's philosophy. In all his works,
\citep{1922ZtesPhy...10..377F,friedman1923mir,1924ZPhy...21..326F},
the physicist from Petrograd insisted that the form of
the Riemannian metric does not resolve this problem. His guide at first was
algebraic topology. Inspired by Poincare's theory of the coverings of the
Riemannian manifolds, he imagined the possibility of a spherical shaped Universe
yet one with infinite diameter and volume.  However, the 3-dimensional
hypersphere $S^{3}$ admits only trivial covering and thus this scenario
seems impossible. Unabashed, Friedman discussed the  ``ramified covering'' of
the sphere, suggesting the ``longitude'' coordinate $\phi$ may run not from
$0$ to $2\pi$ but wind over and over again until infinity. However,
this idea has an obvious flaw: the two poles would be ``covered'' only by one
point and thus the space would no longer be homogeneous (Fig. \ref{belenkiyfig07}).
Within a year, in his quest for the infinite Universe, Friedman found
another argument -- this time a geometrical one.

On advice from his long-term friend and fellow mathematician, Yakov
Tamarkin, Friedman checked whether GR allows solutions for a hyperbolic metric
with a negative space curvature. Indeed, \cite{1924ZPhy...21..326F}
provides a positive answer, with both static and non-static scenarios. 

The static scenario, with constant radius $R$, necessitated zero density
$\rho$ and cosmological constant $\Lambda$ analogous to de Sitter's
solution for the positive curvature case Eq. \ref{belenkiyeq4}. The non-static
scenario, with a time-dependent radius of space-curvature $R=R(t)$, through the
same reasoning as in the positive curvature case, yields the
fundamental equation,
\begin{equation}\label{belenkiyeq11}
\frac{1}{c^{2} }
\dot{R}^{2} =\frac{A+R+\frac{\Lambda }{3c^{2} } R^{3} }{R} ,
\end{equation}
where the integration constant A is related to the average
density $\rho$ as in Eq. \ref{belenkiyeq6}, thus eliminating the possibility
to choose the right curvature sign empirically measuring only density. Besides,
since the volume of the hyperboloid is infinite, the mass of the matter here is
infinite. However, differing from \ref{belenkiyeq5} only in the sign before
the linear term in the cubic, solution \ref{belenkiyeq11} clarified the
ontological significance of the remaining three coefficients in the cubic, which
are the cosmological constant, the sign of the space curvature and the
gravitational radius of the Universe. 

Taking in Eq. \ref{belenkiyeq11} $A=0$ or, equivalently, $\rho=0$ (de Sitter's case),
this solution simplifies to
$R=\varsigma \japblank \sinh ({c\japblank t}/{\varsigma})$,
where in this case
$\varsigma =c\japblank \sqrt{3/\Lambda}$ has the meaning of a scaling parameter. 

The 1924 paper was also ignored. Certainly, Einstein paid no attention
to it. On meeting Lema\^{i}tre in 1927, Einstein called the idea of an expanding
Universe ``abominable.'' But growing astronomical evidence, and most notably
observations by \cite{1929PNAS...15..168H}
and proof by \cite{1930MNRAS..90..668E} that ``solution A'' was unstable,
changed Einstein's mind, though in a somewhat unexpected way.

In 1931 Einstein
recognized Friedman's achievement and suggested purging from GR his old ``nemesis''  -- the
cosmological constant $\Lambda$. The next year, in a joint work,
\cite{1932PNAS...18..213E} promoted an idea of the ``flat''
Universe (i.e., with zero spatial curvature), which is just Eq. \ref{belenkiyeq11}
with $0$ term instead of $+R$ in the upper right side. The latter idea
still exists, together with Friedman's positive
\citep{1922ZtesPhy...10..377F} and negative \citep{1924ZPhy...21..326F}
curvature cases, as neither was preferred by the latest empirical data.  The
former suggestion is not supported by recent astronomical observations: 
$\Lambda$ is still necessary for cosmology.

\section{On Friedman's track: Contributions of Georges Lema\^{i}tre and Edwin
Hubble in the 1920s}

Between Friedman's papers of 1922 and 1924 and the astronomical
observations of the 1990s that led to the 2011 Nobel Prize in Physics there lie
a few groundbreaking achievements: the ``Hubble constant'' that describes the
rate of the Universe's expansion, and the concept of the ``dark matter.''
The \cite{1926ApJ....64..321H} estimate of distances to distant galaxies, led 
\cite{1927ASSB...47...49L} to the discovery of the ``Hubble constant.''
\cite{1934PNAS...20...12L} first gave Friedman's singularity a physical meaning,
that of a ``primeval atom'' that ``blew up'' -- the idea Fred Hoyle described
later as the ``Big Bang.'' Though Einstein did not see any need of the
cosmological constant,
Lema\^{i}tre always valiantly defended its necessity. It is not obvious whose
contribution was decisive in shaping modern cosmology. 

In a popular exposition of GR, \emph{The Meaning of Relativity}, in
three consecutive editions,
\cite{einstein1946meaning,einstein1950meaning,einstein1951meaning},
that included Appendix 1 ``On Cosmological Problem,'' discussing the
unclear nature of the cosmological
constant, Einstein emphasized: ``The mathematician Friedman found a way out of
this dilemma. His result then found a surprising confirmation in Hubble's
discovery of the expansion of the stellar system (a redshift of the spectral
lines which increases uniformly with distance).  The following is essentially
nothing but an exposition of Friedman's idea\dots'' and detailed Friedman's
contribution for the next 15 pages.\footnote{Somewhat unfortunately, Einstein
attributed to Hubble alone what properly belongs to several people, primarily
Slipher and de Sitter. Interestingly, Einstein never quoted any of Lema\^{i}tre's
papers.} 

Later physicists bestowed the title of ``father of modern cosmology''
upon Georges Lema\^{i}tre \citep[][p. 8]{1971phco.book.....P}, while modern
astronomers casually credit Edwin Hubble \citep{2003PhT....56d..53P}.
In the last decade historians decided to clarify this
\citep{2003HisSc..41..141K}. The ``priority'' debate was
narrowed to Lema\^{i}tre and Hubble and concentrated around the passage in 
\cite{1927ASSB...47...49L} which included the
derivation of the ``Hubble constant'' that was omitted in the
English translation of his paper 
\citep{1931MNRAS..91..483L}. Some felt behind this omission stood none other than
Hubble himself - a view which lacking factual support was recently rejected
\citep{2011Natur.479..171L}. Yet, the consensus was that the ``Hubble
constant'' was solely Lema\^{i}tre's idea, thus, supporting Lema\^{i}tre's
right to be called the ``discoverer of the expanding Universe''
\citep[][p. 133]{2009deu..book.....N}.
However we feel the decision was made too quickly. 

\begin{figure}[ht]
\centering
\includegraphics[scale=0.4]{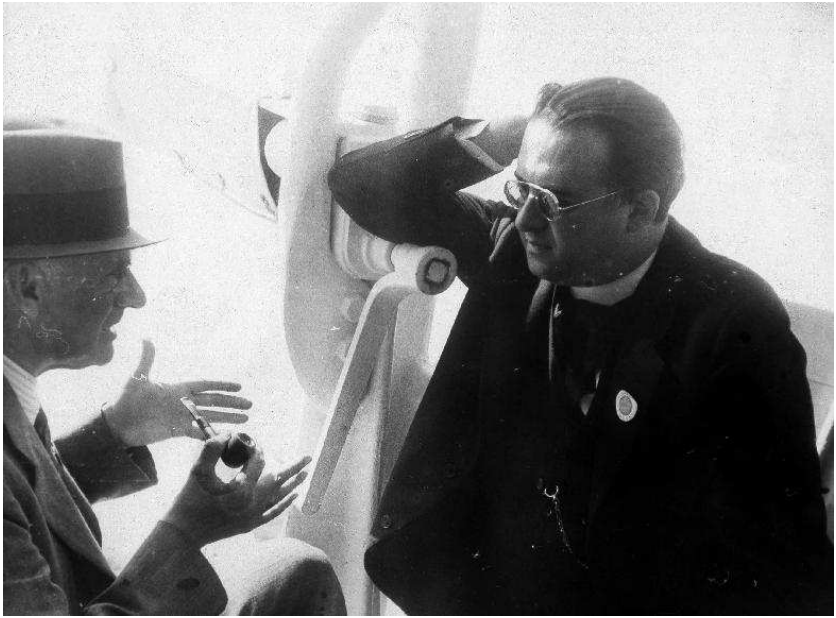}
\caption{Georges Lema\^{i}tre with Arthur Eddington, Stockholm 1938. Courtesy
Georges Lema\^{i}tre Archives,
Catholique University of Louvain.}\label{belenkiyfig08}
\end{figure}

Indeed, though unaware of Friedman's 1922 and 1924 groundbreaking
works, Lema\^{i}tre appeared at the junction when the shortcomings of both
``Solution A'' and ``Solution B'' became pronounced based on the quickly
accumulating astronomical data, coming from Mount Wilson Observatory.
Unlike Friedman, who in 1920-25 for the most part studied mostly meteorology,
Lema\^{i}tre in 1924-1925 was working at MIT on his doctorate dedicated
to cosmology. As a part of Lema\^{i}tre's duties he toured the Mount Wilson
and Lowell Observatories and attended the meetings of various
astronomical societies.  Unlike
Friedman, Lema\^{i}tre had at his desk at least two books with a systematic
exposition of GR and its application to cosmology by
\cite{1923mtr..book.....E} and \cite{silberstein1924theory}, as
well as distances to the galaxies found by
\cite{1926ApJ....64..321H}. Besides, \cite{1923mtr..book.....E}
brought Lema\^{i}tre's attention to the spectral redshift discovery by Slipher,
publishing the ``radial velocities'' of 41 spiral nebulae (Fig. \ref{belenkiyfig02}).

Noticing that the curvature radius $R(t)$ of his metric is
related to the redshift as
$z=\dot{R}dt/R$
and interpreting $dt$ as $r/c$, where $r$ is the distance to a galaxy,
on the one side, and using the
Doppler effect formula $z\approx v/c$, which holds for small ``radial velocity''
$v$, on the other, Lema\^{i}tre was confronted with the
equation $\dot{R}/R=v/r$. This makes sense for the exponentially growing
radius $R(t)$ he found only if $v/r$ is close to a constant.
Seeing a rather weak correlation between $v$ and $r$ for the set of 42 spiral
galaxies,\footnote{\cite{1927ASSB...47...49L}
cites \cite{Stromberg1925ApJ....61..353S}, who made some
``adaptation'' of 41 Slipher's values and added one extra.} cited by several
astronomers, and thus weak factual support for this assumption,
\cite{1927ASSB...47...49L} postulated a \emph{linear relation} between $v$
and $r$. Moreover, he found the coefficient proportionality
$L_{1}=575$ km s$^{-1}$ Mpc$^{-1}$ via a ``simple'' regression method, and
$L_{2}=625$ km s$^{-1}$ Mpc$^{-1}$, via ``weighted'' regression, with
weight $(1+r^{2})^{-1/2}$ attached to each distance $r$ and
corresponding radial velocity $v$. 

The ``simple'' regression method consists of drawing a line from the
beginning of the coordinates to the ``center of gravity'' -- the point with two
coordinates: ``average velocity'' $\overline{v}$ and ``average
distance'' $\overline{r}$.\footnote{This method, a precursor of the OLS
(ordinary least squares), was first invented by Isaac Newton in 1700
\citep{2005AN....326..645B}.} Surprisingly, the same method
was also employed by  \cite{1929PNAS...15..168H} and
\cite{1930BAN.....5..157D}. The latter two, however, tested the linear relation
between $v$ and $r$ from a different perspective -- from the
relation $v/c\approx r/R$, derived by
\cite{silberstein1924theory} from de Sitter's ``solution B'' and popularized by
\cite{1924MNRAS..84..747L}. The constancy of radius $R$ then
immediately leads to $v\approx H\japblank r$. 

\begin{figure}[ht]
\centering
\includegraphics[scale=0.7]{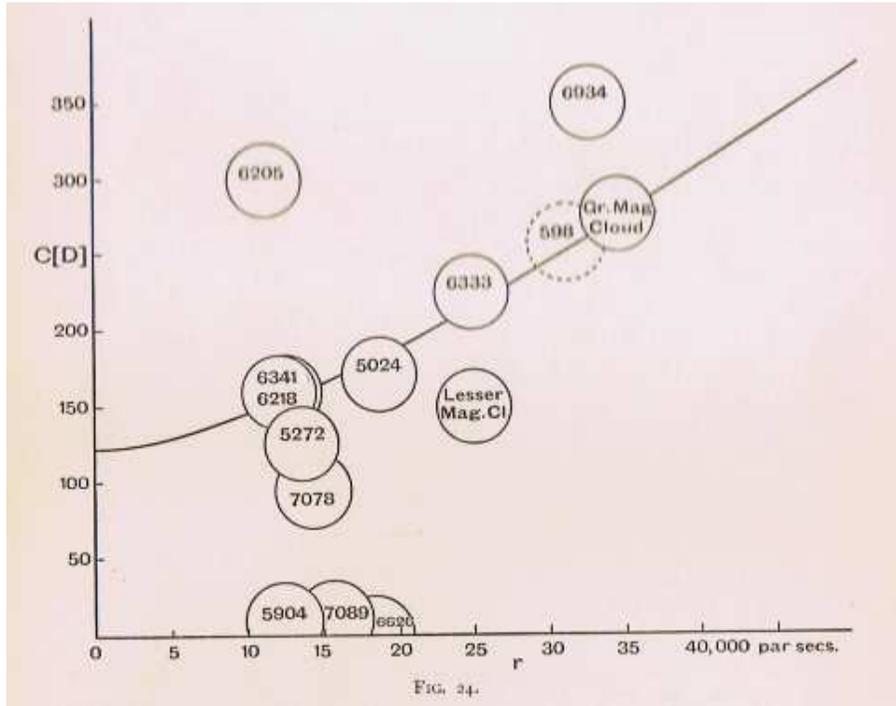}
\caption{The first try to fit the absolute value of the ``red
shift" (the ``Doppler effect'')  $C[D]$ vs. the distance $r$ of 11
globular clusters and two Magellanic Clouds to a semi-linear curve.
From \citet[][p. 551]{silberstein1924theory} with data taken from
Shapley. The dotted circle stands for M33 nebula. All 13 chosen objects are
close to the Sun, lying within 100,000 light years.}\label{belenkiyfig09}
\end{figure}

In postulating the \emph{linear} relation Lema\^{i}tre was influenced by
a similar attempt by \citet[][p. 551]{silberstein1924theory}
to fit Harlow Shapley's data for globular clusters to the asymptotically linear
curve (Fig. \ref{belenkiyfig09}). 

Lema\^{i}tre personally met Silberstein at the meeting of the British
Association of the Advancement of Science in Toronto in August 1924 
\citep{2006AIPC..861.1087F}. There, on August 13, at the ``Cosmical
Physics'' Sub-section, Silberstein gave a talk ``Determination of the Curvature
Radius of Space-Time'' based on de Sitter's model, while Lema\^{i}tre accompanied
Eddington who gave a ``Citizen's Lecture'' on ``Einstein's Theory of
Relativity'' on August 9 \citep[][pp. 19,373]{bhl96033}. In the
paper submitted half a year later, \cite{Lemaitre1925JMP} derives
Silberstein's formula, which linearly connects the radius of the Universe with a
spectral redshift. Silberstein's influence can be seen in the fact that
Lema\^{i}tre finally put more trust in $L_{2}=625$ km s$^{-1}$ Mpc$^{-1}$
obtained via the ``weighted'' regression method, where the weight
$(1+r^{2})^{-1/2}$ attached to distances and velocities is most detrimental to
the more distant galaxies. Silberstein is known to have distrusted the distances
to the ``extra galactic nebulae.'' Pointing out that the different measurements
of the distance to Andromeda nebula, made by Knut Lundmark, were discrepant by
two orders \citep[][p. 522]{silberstein1924theory}, Silberstein based
his computations only on closely lying globular clusters and two Magellanic
Clouds (Fig. \ref{belenkiyfig09}).

The coefficient of proportionality, $L$, could have led to an effective
estimate of the \emph{age} of the Universe ($t_{0}$ in Eq. \ref{belenkiyeq7})
had Lema\^{i}tre embraced a scenario with a \emph{finite}
age of the Universe. However, in his 1927 paper Lema\^{i}tre entirely
ignored the finite age
scenario and the ``Big Bang'' solution in particular. Upon rediscovery of the
Friedman equations (\ref{eq:F1}--\ref{eq:F2}), instead of
considering all classes of solutions, Lema\^{i}tre chose
one particular solution, where the cubic (Fig. \ref{belenkiyfig04}b) has a double
positive root, $x_{0}$. This necessitated accepting a particular (``critical'')
value of $\Lambda _{cr} ={4c^{2} }/{9A^{2}}$ (in Friedman's
notation). Lema\^{i}tre identified point $x_{0} $ with the non-zero initial radius of the
Universe, $R_{E}$. This led Lema\^{i}tre to Friedman's  first ``limiting''
case (Fig. \ref{belenkiyfig06}), where the Universe began
expanding from radius $R_{E}$ to infinity. Thus \cite{1927ASSB...47...49L}
missed the solution with singularity at the origin, M1 World, now
known as the ``Big Bang'' scenario, the most probable scenario for the expanding
Universe according to the astronomical results of 1998-2004 that led to the 2011
Nobel Prize in Physics. 

Even in his 1930 letter to Eddington, Lema\^{i}tre writes: ``I consider a
Universe of curvature constant in space but increasing with time and I emphasize
the existence of a solution in which the motion of the nebulae is always a
receding one from time minus infinity to plus infinity''
\citep[][p. 122]{2009deu..book.....N}. Thus, as late as 1930 Lema\^{i}tre still
adhered to the ``limiting case'' scenario. He abandoned it the following year
after Eddington pointed out that ``such logarithmic infinities have no real
physical significance'' \citep{1931MNRAS..91..490L}, where he
discusses the M2 scenario.  

Only in late 1931, in a short letter to \emph{Nature}, responding to
Eddington, did \cite{1931Natur.127..706L} for the first time consider
the idea of matter coming from a ``discrete number of quanta'' -- a precursor of
the ``Big Bang'' scenario.  Yet by 1931 Lema\^{i}tre had been fully
aware of \cite{1922ZtesPhy...10..377F}  for at least four years, since
his talk with Einstein at the 1927 Solvay conference
\citep[][pp. 111-3]{2009deu..book.....N}.

It is rather surprising that, though recognizing that in his 1929
lecture notes\newline
Lema\^{i}tre thanked Einstein for directing his attention to
Friedman's works, which ``contained several notions and results later
rediscovered by himself,'' \citet[][p. 111]{2009deu..book.....N} conclude:
``Thus, Lema\^{i}tre owes nothing to Friedman.''
The conclusion is acceptable if taken to mean that
\cite{1927ASSB...47...49L} is \emph{independent} of
\cite{1922ZtesPhy...10..377F}. But it is remarkably limited in general if we
compare the contributions of the two and put it in a 1929 perspective. By 1929
Lema\^{i}tre had learned from Friedman the idea of ``birth from singularity'' (the
essence of the ``Big Bang'' scenario) and the idea that the cosmological
constant is a fully independent parameter. Oddly enough, Lema\^{i}tre never
discussed the negative curvature or flat cases.

Lema\^{i}tre had his ``15 months of fame'' in 1930-1931 
\citep[][pp. 126-8]{2009deu..book.....N}, before Einstein was fully
``converted'' to the idea of an expanding Universe in April 1931. Indeed,
already in the first paper written after his trip to the Mount Wilson
Observatory in January 1931 (ibid, pp. 146-7),
\cite{Einstein1931SAW} immediately emphasized Friedman's priority: ``Several
investigators have tried to cope with these new facts by using a spherical space
whose radius is variable in time. The first one who attempted this, uninfluenced
by observations, was A. Friedmann.''\footnote{Translated from the
German by H. Nussbaumer in private communication of September 2012; note that
\citet[][p. 148]{2009deu..book.....N} cite only the first few words.}
Certainly, Einstein admired Friedman for discovering dynamic
solutions purely theoretically, without being driven by the ``facts,'' as
Einstein himself often had been.

\begin{figure}[ht]
\centering
\includegraphics[scale=0.7]{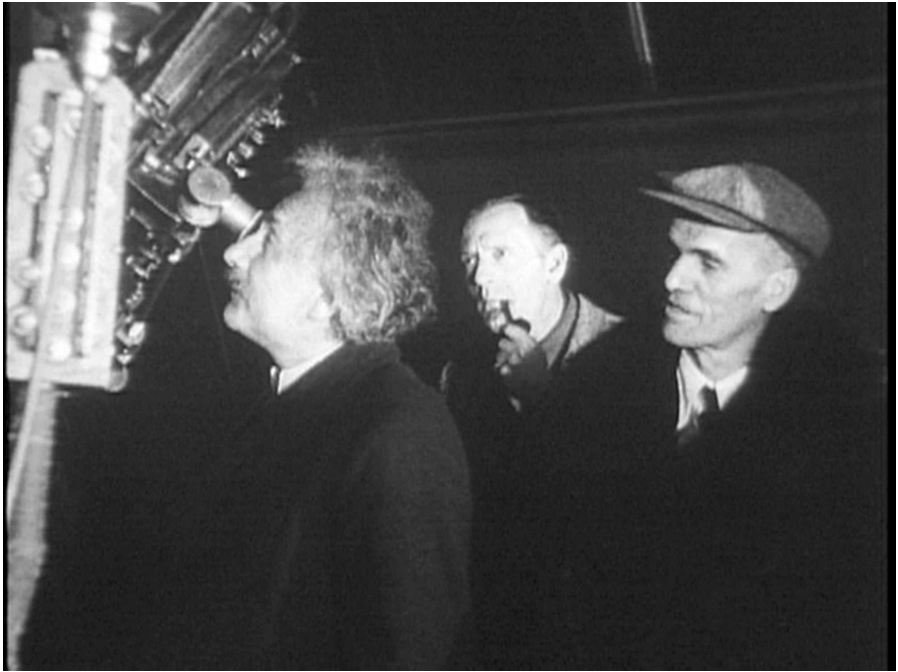}
\caption{Einstein at the 100-inch telescope at the Mount
Wilson Observatory, January 1931, with Edwin Hubble (with a pipe) and Walter
Adams watching. Courtesy of the Archives, California Institute of
Technology.}\label{belenkiyfig10}
\end{figure}

Thereafter Lema\^{i}tre was quoted either only after Friedman or, for
example by Einstein, not quoted at all. Lema\^{i}tre himself recognized the
secondary value of his 1927 work, when in the footnote (2) to
the 1931 English translation of his 1927 paper
\cite{1931MNRAS..91..483L} wrote: ``Equations of the Universe of variable radius
and constant mass have been fully discussed, without reference to the receding
velocity of nebulae, by A. Friedmann (1922).'' Thus, in 1931, Lema\^{i}tre
credited himself only with finding a link between exponentially increasing
radius $R$ of the Universe and redshift phenomena observed for distant galaxies.

Obviously, from a mathematical viewpoint, the Hubble constant
$H_{o}$ is a matter of convenience rather than of fundamental importance,
while Friedman's cubic is fundamental. Deriving three possible scenarios of the
Universe's evolution, Friedman did not need $H_{o}$. In hindsight he was
even fortunate to neglect it. Hubble's grave mistake in
evaluating distances to remote galaxies led to an overestimation of $H_{o}$
and subsequent underestimation of the age of the Universe by a factor of 8 and
thus delayed the acceptance of the ``Big Bang'' scenario by several decades.
Even Einstein in his last years despaired of finding a way out of the dilemma
between a short cosmological age of the Universe, of 1.7 billion years, and a
longer geological age of the Earth! Only at the time of Einstein's death in 1955
did Walter \cite{Baade1952IAU} and Allan \cite{1958ApJ...127..513S}
discover Hubble's mistake in evaluating distances and thus restored
confidence in GR and Friedman's models.

It is interesting how both Friedman and Lema\^{i}tre thought of their
discoveries. Friedman sent both of his papers to the central German physics
journal of his time and used to say privately that he managed to ``horseshoe
Einstein'' \citep{2006alfr.book.....T}. Lema\^{i}tre, in
contrast, not only published his 1927 paper in a little known journal but seemed
to place a low value on his discovery. Quoting Lema\^{i}tre's lecture notes
of 1929, \citet[][p. 111]{2009deu..book.....N} forgot to mention
that in the 23-page long transcript Lema\^{i}tre discusses the astronomical data
only in the context of de Sitter's model, not mentioning at all his model of
1927. ``Lema\^{i}tre mentions that there is a relation between the
velocity of
recession and the distance. Yet he does not provide any number to connect the
velocity of recession with the distance -- the number he himself derived in 1927.
In other words, although the paper contains many numbers, there is no estimate
of the Hubble constant'' \citep{2011arXiv1107.0442S}. Could Lema\^{i}tre
have been that depressed by Einstein's cool reception?

\section{Friedman confirmed: The Nobel Prize 2011 in Physics}

At the end of his 1923 book Friedman concludes:

\begin{quote}
Einstein's theory is justified by experience; it explains the old,
seemingly inexplicable, phenomena and predicts new remarkable relations. The
truest and deepest method to study the world geometry and the structure of the
Universe with the help of Einstein's theory lies in application of this theory
to the whole world and use of astronomical observations. So far this method has
not given us much since mathematical analysis gives up before the difficulties
of the problem and astronomical observations do not provide a reliable basis for
experimental study of the Universe. But these obstacles certainly are of
temporary nature and our descendants, without any doubt, will discover the
structure of the Universe where we are doomed to live.\footnote{Translated
from \citet[][p. 322]{friedman1966works}.}
\end{quote}

Interestingly, the model Friedman seemed to be emotionally attached to
was the ``Periodic World.'' Such a world allows multiple ``births'' and
``deaths'' of the universe - somewhat in tune with the Pythagoras-Plato-Hindu
philosophy of reincarnation. Although already since the late 1980s voices were
heard in favor of the positive cosmological constant
\citep{1990Natur.348..705E}, the basic hypothesis with which astronomers
attacked the problem in the 1990s was in line with
\cite{1932PNAS...18..213E}: that the Universe is flat with zero cosmological
constant and thus -- despite expanding asymptotically as
$R(t)\approx t^{2/3}$ -- decelerates.  

The result was most surprising! Two groups of astronomers, working
independently though using the same technique of observing the so-called
``supernovae Ia,'' found in 1998-1999 that the galaxies at a distance of 4-5
Gly are farther away than they should be according to a constant speed of
expansion given by the currently accepted Hubble constant,
$H_{0} = 70$ km s$^{-1}$ Mpc$^{-1}$ = 2$\times$10$^{-18}$ s$^{-1}$.
Thus, in the last 4-5 billion years,
the Universe has been accelerating driven by the strictly positive cosmological
term \citep{1998AJ....116.1009R,1999ApJ...517..565P}. 

This result however could not discriminate between M1 and M2
worlds since both models allow for acceleration. From early calculations of CMB
anisotropies and their own observations of supernovae, an important relation between
two basic parameters, generalized density $\rho$ (including density of the
``dark matter'') and ``dark energy'' (represented by $\Lambda$) the two groups found
\citep[ibid;][]{2003PhT....56d..53P}:
\begin{equation}\label{belenkiyeq12}
\frac{\Omega _{M}}{\Omega _{\Lambda}} = 
\frac{\kappa c^{2} \rho}{\Lambda} = 
\frac{0.3}{0.7} .
\end{equation}

This result has several applications. On one side, for a somewhat
arbitrary value of the present day average density
$\rho_{0} =3\times10^{-30}$ g cm$^{-3}$,
it leads to $\Lambda =12\times10^{-36} s^{-2}$
(to have $\sqrt{\Lambda /3}=H_{0}$).

On the other side, Eq. \ref{belenkiyeq6} for the same density
$\rho_{0}$ and the present radius of the Universe $R_{0} =1.4\times10^{28}$ cm
leads to $3A/2=8.23\times10^{27}$ cm and gives critical
$\Lambda _{cr} = 4c^{2}/9A^{2} =13\times10^{-36}$ s$^{-2}$,
which is of the same order of magnitude as $\Lambda$.
Thus, Eq. \ref{belenkiyeq12} alone could not be decisive when choosing
between the M1 and M2 worlds. 

The \emph{litmus} test for the choice between M1 and M2 worlds
became the inflexion point. Together with Eq. \ref{belenkiyeq12}, Friedman's
formula (Eq. \ref{belenkiyeq8}) leads to a simple relation between the radius
$R_{f}$ at the inflexion point and the present radius $R_{0}$
\begin{equation}\label{belenkiyeq13}
R_{f} =\sqrt[{3}]{\frac{0.3}{2\japblank 0.7} } R_{0} =0.6\japblank R_{0} .
\end{equation}
Taking $R_{0}$ = 13.75 Gly we get $R_{f}$ = 8.25 Gly.  Thus, one has
to look for galaxies distant from us by more than $r$ = 5.5 Gly. 

The 1998 data are somewhat contradictory on whether the two most
distant supernovae are proportionally fainter than their closer counterparts
(cf. Figs 3 and 4 in \citep{2003PhT....56d..53P}).
Only in 2004 was the question settled \citep{2004ApJ...607..665R}.
According to the formula
for the spectral redshift $z$, given as a result of coupling
relativistic Doppler and gravitational effects
\citep{1961cosm.book.....B}:
\begin{equation} \label{belenkiyeq14}
1+z=\sqrt{\frac{1+{\raise0.7ex\hbox{$ v $}\!\mathord{\left/{\vphantom{v
c}}\right.\kern-\nulldelimiterspace}\!\lower0.7ex\hbox{$ c $}}
}{1-{\raise0.7ex\hbox{$ v $}\!\mathord{\left/{\vphantom{v
c}}\right.\kern-\nulldelimiterspace}\!\lower0.7ex\hbox{$ c $}} } \japblank } \,
\left(1-\frac{\kappa \rho _{0} r^{2} }{12} \right)     ,
\end{equation}
the galaxies observed at $r=8.4-10.5$ Gly away ($1<z<1.6$), are
closer to the Milky Way than they should be, given the acceleration rate shown
by the galaxies 5 Gly away. Thus, at some time, the Universe switched from
deceleration to acceleration. The inflexion point was estimated as being at
about $R_{f} =5\pm 1$Gly ($z=0.46\pm 0.13$), and the M1 model triumphed.

It is interesting that for positive space curvature, and thus a finite
mass Universe, the inflexion point is close or even identical to the
Schwarzschild radius of the Universe, $R_{S} =2GM/c^{2} $, which (for the
finite-mass- and positive-curvature-Universe) for the same density $\rho_{0}$
and radius $R_{0}$ as above can be found as
\begin{equation} \label{belenkiyeq15}
R_{S} =\frac{2G\japblank M}{c^{2} } =\frac{\kappa \japblank \rho
_{0} \japblank R_{0}^{3} }{2} \approx 0.6\japblank R_{0} .
\end{equation}
If for some reason these two points are identical, the inflexion
point marks the point when the Universe turned from being a ``black hole'' to
something visible to the other universes.

\section{Friedman's Legacy: Ninety Years Later}

Though the question of the physical scenario of the ``Big Bang,''
first raised by \cite{1934PNAS...20...12L} and expounded by
George \cite{gamow1952creation} and others, is not yet settled, the mathematical
(kinematic) part of the ``Big Bang,'' described by
\cite{1922ZtesPhy...10..377F,1924ZPhy...21..326F}, largely is.

There is a tendency, based on one remark by Vladimir Fock
\citep[][p. 402]{friedman1966works}, to present Friedman as merely a
mathematician, unconcerned with the physical implications of his discovery
\citep{2003HisSc..41..141K}. This remark and its implications,
however, completely contradict Friedman's image as a mathematical physicist with
serious achievements in meteorology and hydrodynamics. The wide spectrum of the
problems he solved, as seen in Friedman's ``Collected works''
\citep[][pp. 424-447]{friedman1966works}, leaves no doubt that he was interested in
verification of his theories.  Sadly, Friedman's premature death in 1925 did not
allow him to fulfill this goal for the cosmology problem. 

Another argument for why Friedman does not deserve the title of
discoverer of the expanding Universe is that he ``failed to provide a physical
reason for the expanding Universe.''\footnote{R. Smith, private communication
of Sept. 15, 2012.}
The simple answer to this allegation is that Friedman believed, as many of us
still do, that Einstein's GR equations correctly describe physical reality.  

The third argument is that Friedman ``did not take part in the debates of 1930s
when the expanding Universe concept was widely
accepted.''\footnote{H. Nussbaumer, private communication of Sept. 15, 2012.}
This is untrue in two
different ways. First, as we have seen, the debates of 1930s came to a dead end
because of the wrong, greatly underestimated Hubble constant. As a result, other
scenarios, like the ``Steady State'' Universe were proposed in 1948
\citep{1961cosm.book.....B}, delaying the recognition of
the ``Big Bang'' scenario. On the other hand, the \cite{1932PNAS...18..213E} paper,
which set the standard for the ``Big Bang'' cosmology for many decades until 1998,
was impossible without 
\cite{1922ZtesPhy...10..377F,1924ZPhy...21..326F}, where
the first paper provided the theoretical basis for elimination of the
cosmological constant, while the second opened the door for a flat Universe. 

As a result of these ``arguments,'' Friedman's pioneering role in
modern cosmology is often underestimated or misrepresented. For a long time both
of Friedman's papers were quoted by various authors without even
mentioning him in the body of the
book.\footnote{An outstanding example is given by \cite{1961cosm.book.....B}.}
A more recent example, Friedman's
photo is not present on the cover of the quite comprehensive book
\emph{Discovering the Expanding Universe} by \cite{2009deu..book.....N} 
although all other founding fathers of modern cosmology are there. 

It is clear, however, that in shaping the theoretical part of modern
cosmology \cite{1922ZtesPhy...10..377F} went much further than his
predecessors and even immediate successors, like \cite{1927ASSB...47...49L}.
According to
the memoirs of his wife, Ekaterina Friedman, on this and other occasions her
husband used to say citing Dante: ``The waters I am entering, no one yet has
crossed!''\citep[][p. 396]{friedman1966works}.
And indeed, Friedman's approach was the first correct application of
GR to cosmology, which brought forward the idea of the \emph{expanding} Universe,
possibly born from a singularity. Moreover, realizing that GR may admit
different metrics, \cite{1924ZPhy...21..326F} alerted physicists
that the Universe could be infinite and negatively curved.

As a philosopher of cosmology, Friedman stands head and
shoulders above all the other participants of the
great cosmological debate in the 1920s, including Einstein. It is a well-known
fact that Einstein later bemoaned the introduction of the cosmological constant
since the expansion of the Universe in Friedman's models could be
achieved with a \emph{zero} cosmological constant. Only in 1930 did Eddington and
de Sitter embrace the expanding Universe scenario, with the latter
admitting that the ``veil fell from his eyes.''
Only in 1931 did Lema\^{i}tre begin thinking of alternatives to his
initial model of ``logarithmically long awakening'' of the Universe from the
non-zero radius. In contrast, it is known that Hubble never embraced the
expanding Universe model \citep[][p. 120]{2009deu..book.....N}.

Recognizing the scale of his achievements, in the last three editions
of the \emph{Meaning of Relativity}, Einstein acknowledged Friedman's
theoretical groundwork together with Hubble's astronomical observations, viewing
contributions of others as secondary. Sadly, Einstein's words were ignored.
After the 1930s, Hubble alone became credited for the ``Hubble constant,'' while
Lema\^{i}tre received all the credit for the ``Big Bang'' theory. However, the
persistence of Soviet physicists, who, since the early 1960s, when Soviet
communist rulers stopped condemning ``Lema\^{i}tre's reactionary theory,'' and the
``Big Bang'' theory received a fresh breath, began raising their voices on
behalf of Friedman's achievements, finally paid off and since the 1980s
Friedman's metric and equations began to carry his name.\footnote{Remarkably,
it appears that the first who used the expression ``Friedmann's equation"
(with respect to Eq. 7) was \cite{1949RvMP...21..357L}.}

The staunchest defender of Friedman's legacy, Yakov Zel'dovich, pointed
to a singularly touching detail: ``Friedman published his works in 1922-1924, in
a time of great hardships. Herbert Wells' impression about Moscow and
Petrograd of 1921 was of `Russia [lying] in the Shadows.'
In the same issue of the 1922 journal where Friedman's paper appeared, there was
an appeal to German scientists to donate scientific literature to their Soviet
colleagues who had been separated from it during the war and the revolution.
Under those circumstances, Friedman's daring discovery was not only a scientific
but also a human feat!''\footnote{Translated from \citet[][p. 404]{friedman1966works}.}

\begin{figure}[ht]
\centering
\includegraphics[scale=0.54]{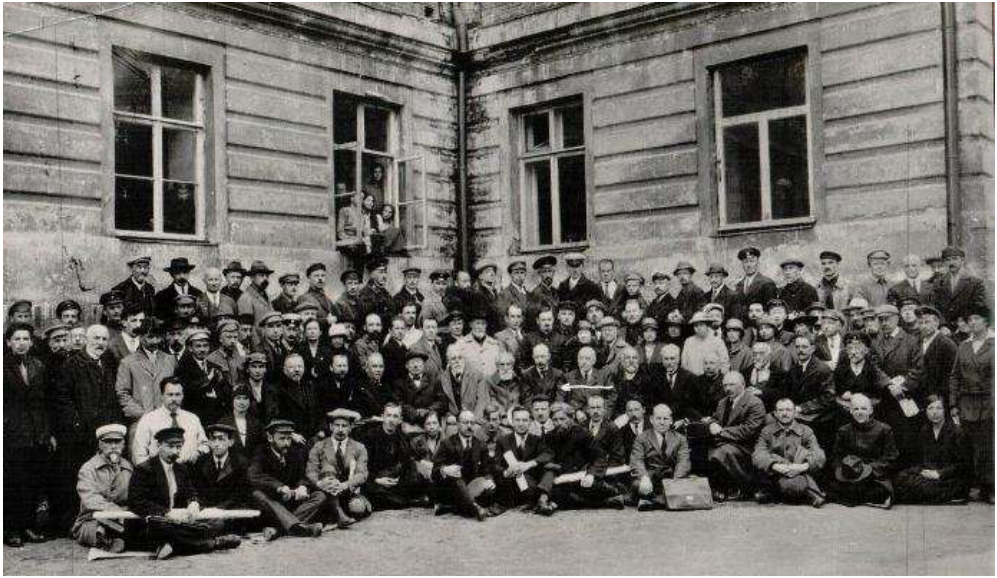}
\caption{Participants of the Third Meteorological Congress,
Moscow, May 1925 held at the 1$^{st}$ Moscow State University. Friedman
sits in the second row in the center (see arrow). Courtesy of
the Voeikov Main Geophysical Observatory (St-Petersburg).}\label{belenkiyfig11}
\end{figure}

And indeed, Friedman's letter to Ehrenfest on June 3, 1922
(Fig. \ref{belenkiyfig03}), announcing the
discovery of expanding universes, reveals the level of deprivation of
Russian society at that time. In the letter, he asks to be mailed an
off-print of de Sitter's 1917 paper from the \emph{Monthly Notices of the Royal
Astronomical Society}, saying ``although the journal might be found in Pulkovo, a
suburb of Petrograd, there is no way to fetch it from there and going by foot
would be extremely difficult.''  

True, in the three subsequent years remaining to him Friedman's life
conditions gradually improved. Alexander Friedman quickly ascended
the academic and administrative ladders (Fig. \ref{belenkiyfig11})
and his early death seems as unfair as
in the cases of two other founding fathers of modern cosmology:
Hermann Minkowski and Karl Schwarzschild.
In Friedman's obituary, Vladimir Steklov enumerates a variety of papers
and books Friedman published during these three years,\footnote{Friedman's
bibliography includes 25 titles published between 1922 and 1925. All 25
come after his first paper on cosmology and are exactly half of all of
his publications \citep[][pp. 456-7]{friedman1966works}.}
and cites the letter from Professor Heinrich von Ficker, Director of the
Prussian Meteorological Institute, which says in particular that
``with Friedman's death you lost one of the most remarkable disciples,
one who will be mourned by every meteorologist. The strongest hope of
theoretical meteorology departed with him. This case is especially sad
for me since among Russian meteorologists he was the closest to me."
Steklov concludes: ``A.A. Friedman died, 37 years old, in the peak of
strength and talent." 

One may wonder what Friedman could have accomplished
had he be given several more years -- at least until 1937.
And how much brighter would Friedman's scientific star shine had
Einstein not been initially blind to his revolutionary discoveries.

\acknowledgements

The author acknowledges helpful information about Friedman from Carlo Beenakker
(Leiden University), V.~M. Kattsov
and E.~L. Makhotkina (Voeikov Main Geophysical Observatory, St-Petersburg),
Sabine Lehr (Springer DE), Liliane
Moens-Haulotte (Georges Lema\^{i}tre Centre for Earth and Climate Research),
as well as discussions with Alexei Kojevnikov (UBC), Harry Nussbaumer
(ETH Zurich), John Peacock (Edinburgh University), Todd Timberlake
(Berry College), Michael Way (NASA) and Sarah Olesh (Vancouver).

\bibliography{belenkiy}

\begin{thebibliography}{}
\expandafter\ifx\csname natexlab\endcsname\relax\def\natexlab#1{#1}\fi
\expandafter\ifx\csname url\endcsname\relax
  \def\url#1{\texttt{#1}}\fi
\expandafter\ifx\csname urlprefix\endcsname\relax\def\urlprefix{URL }\fi
\providecommand{\eprint}[2][]{\url{#2}}

\bibitem[{{Baade}(1952)}]{Baade1952IAU}
{Baade}, W. 1952, {A Revision of the Extra-Galactic Distance Scale},
  Transactions of the International Astronomical Union, 8, 397

\bibitem[{BAAS(1925)}]{bhl96033}
BAAS 1925, Report of the British Association for the Advancement of Science.,
  vol. 92nd Meeting (1924) (London.).
  Http://www.biodiversitylibrary.org/bibliography/2276,
  \urlprefix\url{http://www.biodiversitylibrary.org/item/96033}

\bibitem[{{Beenakker}(2012)}]{Beenakker2012}
{Beenakker}, C. 2012, {Friedmann Papers} (Lorentz-Institute, Leiden
  University). \urlprefix\url{http://www.lorentz.leidenuniv.nl/Friedmann}

\bibitem[{Belenkiy(2012)}]{belenkiy:38}
Belenkiy, A. 2012, {Alexander Friedmann and the Origins of Modern Cosmology},
  Physics Today, 65, 38. \urlprefix\url{http://link.aip.org/link/?PTO/65/38/1}

\bibitem[{{Belenkiy} \& {Vila-Echag\"{u}e}(2005)}]{2005AN....326..645B}
{Belenkiy}, A., \& {Vila-Echag\"{u}e}, E. 2005, {History of One Defeat: Reform
  of the Julian Calendar as Envisioned by Isaac Newton}, Notes and Records of
  the Royal Society, 59, 223.
  \urlprefix\url{http://rsnr.royalsocietypublishing.org/content/59/3/223.abstract}

\bibitem[{{Bondi}(1961)}]{1961cosm.book.....B}
{Bondi}, H. 1961, {Cosmology} (Cambridge University Press)

\bibitem[{Corry et~al.(1997)Corry, Renn, \& Stachel}]{Corry1997}
Corry, L., Renn, J., \& Stachel, J. 1997, Belated decision in the
  hilbert-einstein priority dispute, Science, 278, 1270.
  \eprint{http://www.sciencemag.org/content/278/5341/1270.full.pdf},
  \urlprefix\url{http://www.sciencemag.org/content/278/5341/1270.abstract}

\bibitem[{{de Sitter}(1917)}]{deSitter1917MNRAS..78....3D}
{de Sitter}, W. 1917, {Einstein's Theory of Gravitation and its Astronomical
  Consequences. Third paper}, \mnras, 78, 3

\bibitem[{{de Sitter}(1930)}]{1930BAN.....5..157D}
--- 1930, {On the Magnitudes, Diameters and Distances of the Extragalactic
  Nebulae and their Apparent Radial Velocities (Errata: 5 V, 230)}, Bulletin of
  the Astronomical Institutes of the Netherlands, 5, 157

\bibitem[{{Eddington}(1920)}]{1920stga.book.....E}
{Eddington}, A.~S. 1920, {Space, Time and Gravitation. An Outline of the
  General Relativity Theory} (Cambridge University Press)

\bibitem[{{Eddington}(1923)}]{1923mtr..book.....E}
--- 1923, {The Mathematical Theory of Relativity} (Cambridge University Press)

\bibitem[{{Eddington}(1930)}]{1930MNRAS..90..668E}
--- 1930, {On the Instability of Einstein's Spherical World}, \mnras, 90, 668

\bibitem[{{Efstathiou} et~al.(1990){Efstathiou}, {Sutherland}, \&
  {Maddox}}]{1990Natur.348..705E}
{Efstathiou}, G., {Sutherland}, W.~J., \& {Maddox}, S.~J. 1990, {The
  Cosmological Constant and Cold Dark Matter}, \nat, 348, 705

\bibitem[{{Einstein}(1917)}]{1917SPAW.......142E}
{Einstein}, A. 1917, {Kosmologische Betrachtungen zur Allgemeinen
  Relativit{\"a}tstheorie}, Sitzungsberichte der K{\"o}niglich Preu{\ss}ischen
  Akademie der Wissenschaften (Berlin), Seite 142-152., 142

\bibitem[{{Einstein}(1922)}]{Einstein1922ZPhy...11..326E}
--- 1922, {Bemerkung zu der Arbeit von A. Friedmann {\"U}ber die Kr{\"u}mmung
  des Raumes}, Zeitschrift fur Physik, 11, 326

\bibitem[{{Einstein}(1923)}]{Einstein1923ZPhy...16..228E}
--- 1923, {Notiz zu der Arbeit von A. Friedmann {\"U}ber die Kr{\"u}mmung des
  Raumes}, Zeitschrift fur Physik, 16, 228

\bibitem[{{Einstein}(1931)}]{Einstein1931SAW}
--- 1931, {Zum kosmologischen Problem der Allgemeinen Relativit\"{a}tstheorie},
  {Sitzungsberichte der Preussischen Akademie der Wissenschaften,
  Physikalisch-mathematische Klasse}, 235

\bibitem[{{Einstein}(1946)}]{einstein1946meaning}
--- 1946, {The Meaning of Relativity: Third Edition with an Appendix}
  (Princeton University Press)

\bibitem[{{Einstein}(1950)}]{einstein1950meaning}
--- 1950, {The Meaning of Relativity: with Further Appendix} (Princeton
  University Press)

\bibitem[{{Einstein}(1951)}]{einstein1951meaning}
--- 1951, {The Meaning of Relativity} (Princeton University Press)

\bibitem[{{Einstein} \& {de Sitter}(1932)}]{1932PNAS...18..213E}
{Einstein}, A., \& {de Sitter}, W. 1932, {On the Relation between the Expansion
  and the Mean Density of the Universe}, Proceedings of the National Academy of
  Science, 18, 213

\bibitem[{{Flin} \& {Duerbeck}(2006)}]{2006AIPC..861.1087F}
{Flin}, P., \& {Duerbeck}, H.~W. 2006, in Albert Einstein Century International
  Conference, edited by J.-M. {Alimi}, \& A.~{F{\"u}zfa}, vol. 861 of American
  Institute of Physics Conference Series, 1087

\bibitem[{{Friedman}(1922)}]{1922ZtesPhy...10..377F}
{Friedman}, A. 1922, {{\"U}ber die Kr{\"u}mmung des Raumes}, Zeitschrift
  f\"{u}r Physik, 10, 377

\bibitem[{{Friedman}(1923)}]{friedman1923mir}
--- 1923, {Mir Kak Prostranstvo i Vremya [The World as Space and Time]}
  (Petrograd: Academia)

\bibitem[{{Friedman}(1924)}]{1924ZPhy...21..326F}
--- 1924, {{\"U}ber die M{\"o}glichkeit einer Welt mit Konstanter Negativer
  Kr{\"u}mmung des Raumes}, Zeitschrift f\"{u}r Physik, 21, 326

\bibitem[{{Friedman}(1966)}]{friedman1966works}
--- 1966, {Collected Works} (Nauka, Moscow)

\bibitem[{{Friedman}(1999{\natexlab{a}})}]{1999GReGr..31.1991F}
--- 1999{\natexlab{a}}, {On the Curvature of Space}, General Relativity and
  Gravitation, 31, 1991

\bibitem[{{Friedman}(1999{\natexlab{b}})}]{1999GReGr..31.2001F}
--- 1999{\natexlab{b}}, {On the Possibility of a World wih Constant Negative
  Curvature of Space}, General Relativity and Gravitation, 31, 2001

\bibitem[{{Gamow}(1952)}]{gamow1952creation}
{Gamow}, G. 1952, The Creation of the Universe (Viking Press).
  \urlprefix\url{http://books.google.com/books?id=orO7AAAAIAAJ}

\bibitem[{{Hubble}(1925)}]{Hubble1925PA.....33..252H}
{Hubble}, E.~P. 1925, {Cepheids in Spiral Nebulae}, Popular Astronomy, 33, 252

\bibitem[{{Hubble}(1926)}]{1926ApJ....64..321H}
--- 1926, {Extragalactic Nebulae.}, \apj, 64, 321

\bibitem[{{Hubble}(1929)}]{1929PNAS...15..168H}
--- 1929, {A Relation between Distance and Radial Velocity among Extra-Galactic
  Nebulae}, Proceedings of the National Academy of Science, 15, 168

\bibitem[{{Jeans}(1928)}]{Jeans1928astronomy}
{Jeans}, J.~H. 1928, Astronomy and Cosmogony (Cambridge University Press).
  \urlprefix\url{http://books.google.se/books?id=Sf8QcgAACAAJ}

\bibitem[{Klein(1928)}]{klein1928vorlesungen}
Klein, F. 1928, Vorlesungen {\"u}ber Nichteuklidische Geometrie, Die
  Grundlehren der mathematischen Wissenschaften (Springer-Verlag).
  \urlprefix\url{http://books.google.ca/books?id=lPFdtwAACAAJ}

\bibitem[{{Kragh} \& {Smith}(2003)}]{2003HisSc..41..141K}
{Kragh}, H., \& {Smith}, R.~W. 2003, {Who Discovered the Expanding Universe?},
  History of Science, 41, 141

\bibitem[{{Lanczos}(1923)}]{1923ZPhy...17..168L}
{Lanczos}, K. 1923, {{\"U}ber die Rotverschiebung in der de Sitterschen Welt},
  Zeitschrift f\"{u}r Physik, 17, 168

\bibitem[{{Lema{\^i}tre}(1925)}]{Lemaitre1925JMP}
{Lema{\^i}tre}, G. 1925, {Note on de Sitter's universe}, {Journal of
  Mathematics and Physics}, 4, 188

\bibitem[{{Lema{\^i}tre}(1927)}]{1927ASSB...47...49L}
--- 1927, {Un Univers Homog{\`e}ne de Masse Constante et de Rayon Croissant
  Rendant Compte de la Vitesse Radiale des N{\'e}buleuses Extra-Galactiques},
  Annales de la Societe Scietifique de Bruxelles, 47, 49

\bibitem[{{Lema{\^i}tre}(1931{\natexlab{a}})}]{1931MNRAS..91..483L}
--- 1931{\natexlab{a}}, {\mockalph{aa}Expansion of the Universe, A Homogeneous
  Universe of Constant Mass and Increasing Radius Accounting for the Radial
  Velocity of Extra-Galactic Nebulae}, \mnras, 91, 483

\bibitem[{{Lema{\^i}tre}(1931{\natexlab{b}})}]{1931MNRAS..91..490L}
--- 1931{\natexlab{b}}, {\mockalph{bb}The Expanding Universe}, \mnras, 91, 490

\bibitem[{{Lema{\^i}tre}(1931{\natexlab{c}})}]{1931Natur.127..706L}
--- 1931{\natexlab{c}}, {\mockalph{cc}The Beginning of the World from the Point
  of View of Quantum Theory.}, \nat, 127, 706

\bibitem[{{Lema{\^i}tre}(1934)}]{1934PNAS...20...12L}
--- 1934, {Evolution of the Expanding Universe}, Proceedings of the National
  Academy of Science, 20, 12

\bibitem[{{Lema{\^i}tre}(1949)}]{1949RvMP...21..357L}
--- 1949, {Cosmological Application of Relativity}, Reviews of Modern Physics,
  21, 357

\bibitem[{{Livio}(2011)}]{2011Natur.479..171L}
{Livio}, M. 2011, {Lost in Translation: Mystery of the Missing Text Solved},
  \nat, 479, 171

\bibitem[{{Lundmark}(1924)}]{1924MNRAS..84..747L}
{Lundmark}, K. 1924, {The Determination of the Curvature of Space-Time in de
  Sitter's world}, \mnras, 84, 747

\bibitem[{Nobelprize.org(2011)}]{nobel2011}
Nobelprize.org 2011, {The Accelerating Universe} (Swedish Academy of Sciences).
  {Scientific Background on the Nobel Prize in Physics 2011: Compiled by the
  Class for Physics of the Royal Swedish Academy of Sciences},
  \urlprefix\url{http://www.nobelprize.org/nobel_prizes/physics/laureates/2011/}

\bibitem[{{Nussbaumer} \& {Bieri}(2009)}]{2009deu..book.....N}
{Nussbaumer}, H., \& {Bieri}, L. 2009, {Discovering the Expanding Universe}
  (Cambridge University Press)

\bibitem[{{Peebles}(1971)}]{1971phco.book.....P}
{Peebles}, P.~J.~E. 1971, {Physical Cosmology} (Princeton University Press)

\bibitem[{{Perlmutter}(2003)}]{2003PhT....56d..53P}
{Perlmutter}, S. 2003, {Supernovae, Dark Energy, and the Accelerating
  Universe}, Physics Today, 56, 040000

\bibitem[{{Perlmutter} et~al.(1999){Perlmutter}, {Aldering}, {Goldhaber},
  {Knop}, {Nugent}, {Castro}, {Deustua}, {Fabbro}, {Goobar}, {Groom}, {Hook},
  {Kim}, {Kim}, {Lee}, {Nunes}, {Pain}, {Pennypacker}, {Quimby}, {Lidman},
  {Ellis}, {Irwin}, {McMahon}, {Ruiz-Lapuente}, {Walton}, {Schaefer}, {Boyle},
  {Filippenko}, {Matheson}, {Fruchter}, {Panagia}, {Newberg}, {Couch}, \&
  {Supernova Cosmology Project}}]{1999ApJ...517..565P}
{Perlmutter}, S., {Aldering}, G., {Goldhaber}, G., {Knop}, R.~A., {Nugent}, P.,
  {Castro}, P.~G., {Deustua}, S., {Fabbro}, S., {Goobar}, A., {Groom}, D.~E.,
  {Hook}, I.~M., {Kim}, A.~G., {Kim}, M.~Y., {Lee}, J.~C., {Nunes}, N.~J.,
  {Pain}, R., {Pennypacker}, C.~R., {Quimby}, R., {Lidman}, C., {Ellis}, R.~S.,
  {Irwin}, M., {McMahon}, R.~G., {Ruiz-Lapuente}, P., {Walton}, N., {Schaefer},
  B., {Boyle}, B.~J., {Filippenko}, A.~V., {Matheson}, T., {Fruchter}, A.~S.,
  {Panagia}, N., {Newberg}, H.~J.~M., {Couch}, W.~J., \& {Supernova Cosmology
  Project} 1999, {Measurements of Omega and Lambda from 42 High-Redshift
  Supernovae}, \apj, 517, 565. \eprint{arXiv:astro-ph/9812133}

\bibitem[{{Riess} et~al.(1998){Riess}, {Filippenko}, {Challis}, {Clocchiatti},
  {Diercks}, {Garnavich}, {Gilliland}, {Hogan}, {Jha}, {Kirshner},
  {Leibundgut}, {Phillips}, {Reiss}, {Schmidt}, {Schommer}, {Smith},
  {Spyromilio}, {Stubbs}, {Suntzeff}, \& {Tonry}}]{1998AJ....116.1009R}
{Riess}, A.~G., {Filippenko}, A.~V., {Challis}, P., {Clocchiatti}, A.,
  {Diercks}, A., {Garnavich}, P.~M., {Gilliland}, R.~L., {Hogan}, C.~J., {Jha},
  S., {Kirshner}, R.~P., {Leibundgut}, B., {Phillips}, M.~M., {Reiss}, D.,
  {Schmidt}, B.~P., {Schommer}, R.~A., {Smith}, R.~C., {Spyromilio}, J.,
  {Stubbs}, C., {Suntzeff}, N.~B., \& {Tonry}, J. 1998, {Observational Evidence
  from Supernovae for an Accelerating Universe and a Cosmological Constant},
  \aj, 116, 1009. \eprint{arXiv:astro-ph/9805201}

\bibitem[{{Riess} et~al.(2004){Riess}, {Strolger}, {Tonry}, {Casertano},
  {Ferguson}, {Mobasher}, {Challis}, {Filippenko}, {Jha}, {Li}, {Chornock},
  {Kirshner}, {Leibundgut}, {Dickinson}, {Livio}, {Giavalisco}, {Steidel},
  {Ben{\'{\i}}tez}, \& {Tsvetanov}}]{2004ApJ...607..665R}
{Riess}, A.~G., {Strolger}, L.~G., {Tonry}, J., {Casertano}, S., {Ferguson},
  H.~C., {Mobasher}, B., {Challis}, P., {Filippenko}, A.~V., {Jha}, S., {Li},
  W., {Chornock}, R., {Kirshner}, R.~P., {Leibundgut}, B., {Dickinson}, M.,
  {Livio}, M., {Giavalisco}, M., {Steidel}, C.~C., {Ben{\'{\i}}tez}, T., \&
  {Tsvetanov}, Z. 2004, {Type Ia Supernova Discoveries at z $>$ 1 from the
  Hubble Space Telescope: Evidence for Past Deceleration and Constraints on
  Dark Energy Evolution}, \apj, 607, 665. \eprint{arXiv:astro-ph/0402512}

\bibitem[{{Sandage}(1958)}]{1958ApJ...127..513S}
{Sandage}, A. 1958, {Current Problems in the Extragalactic Distance Scale.},
  \apj, 127, 513

\bibitem[{{Shaviv}(2011)}]{2011arXiv1107.0442S}
{Shaviv}, G. 2011, {Did Edwin Hubble plagiarize?}, ArXiv e-prints.
  \eprint{1107.0442}

\bibitem[{Silberstein(1924)}]{silberstein1924theory}
Silberstein, L. 1924, The Theory of Relativity (Macmillan).
  \urlprefix\url{http://books.google.ca/books?id=K91FAQAAIAAJ}

\bibitem[{{Slipher}(1913)}]{Slipher1913LowOB...2...56S}
{Slipher}, V.~M. 1913, {The Radial Velocity of the Andromeda Nebula}, Lowell
  Observatory Bulletin, 2, 56

\bibitem[{{Stromberg}(1925)}]{Stromberg1925ApJ....61..353S}
{Stromberg}, G. 1925, {Analysis of Radial Velocities of Globular Clusters and
  Non-Galactic Nebulae.}, \apj, 61, 353

\bibitem[{{Tropp} et~al.(2006){Tropp}, {Frenkel}, {Chernin}, {Dron}, \&
  {Burov}}]{2006alfr.book.....T}
{Tropp}, E.~A., {Frenkel}, V.~Y., {Chernin}, A.~D., {Dron}, A., \& {Burov}, M.
  2006, {Alexander A Friedmann} (Cambridge University Press)

\bibitem[{Weyl(1918)}]{weyl1918raum}
Weyl, H. 1918, Raum. Zeit. Materie: Vorlesungen {\"u}ber Allgemeine
  Relativit{\"a}tstheorie (J. Springer).
  \urlprefix\url{http://books.google.ca/books?id=eiVMAAAAMAAJ}

\bibitem[{Weyl(1923)}]{weyl1923PZ}
--- 1923, {Zur Allgemeinen Relativit\"{a}tstheorie}, Physikalische Zeitschrift,
  24, 230

\end{thebibliography}

\end{document}